\documentclass[twocolumn]{aastex63}

\usepackage{natbib}
\usepackage[style=american]{csquotes}
\newcommand{\tess}{\emph{TESS}}

\usepackage{scrextend}
\let\orgautoref\autoref
\renewcommand{\autoref}
        {\def\equationautorefname{Eq.}%
         \def\figureautorefname{Fig.}%
         \def\sectionautorefname{Sect.}%
         \def\subsectionautorefname{Sect.}%
         \def\subsubsectionautorefname{Sect.}%
         \orgautoref}

\usepackage{changepage} 
\usepackage{rotating}

\received{April 27, 2021}
\accepted{----}

\shorttitle{A pair of warm giant planets near the 2:1 MMR.}
\shortauthors{Trifonov et al.}

\begin{document}

\title{A pair of warm giant planets near the 2:1 mean motion resonance around the K-dwarf star TOI-2202\footnote{Based on observations collected at the European Organization for Astronomical Research in the Southern Hemisphere under ESO programmes 0104.C-0413, 1102.C-0923, and MPG programmes 0102.A-9006,
0103.A-9008,
0104.A-9007. This paper includes data gathered with the 6.5 meter Magellan Telescopes located at Las Campanas Observatory, Chile.}}

\correspondingauthor{Trifon Trifonov}
\email{trifonov@mpia.de}

\author[0000-0002-0236-775X]{Trifon Trifonov}
\affil{Max-Planck-Institut f\"{u}r Astronomie, K\"{o}nigstuhl  17, 69117 Heidelberg, Germany}

\author[0000-0002-9158-7315]{Rafael Brahm}
\affil{Facultad de Ingeniera y Ciencias, Universidad Adolfo Ib\'{a}\~{n}ez, Av. Diagonal las Torres 2640, Pe\~{n}alol\'{e}n, Santiago, Chile}
\affil{Millennium Institute for Astrophysics, Chile}

\author[0000-0001-9513-1449]{Nestor Espinoza}
\affil{Space Telescope Science Institute, 3700 San Martin Drive, Baltimore, MD 21218, USA}

\author[0000-0002-1493-300X]{Thomas Henning}
\affil{Max-Planck-Institut f\"{u}r Astronomie, K\"{o}nigstuhl  17, 69117 Heidelberg, Germany}

\author[0000-0002-5389-3944]{Andr\'es Jord\'an}
\affil{Facultad de Ingeniera y Ciencias, Universidad Adolfo Ib\'{a}\~{n}ez, Av. Diagonal las Torres 2640, Pe\~{n}alol\'{e}n, Santiago, Chile}
\affil{Millennium Institute for Astrophysics, Chile}

\author{David Nesvorny}
\affil{Department of Space Studies, Southwest Research Institute, 1050 Walnut Street, Suite 300, Boulder, CO 80302, USA}

\author[0000-0001-9677-1296]{Rebekah I. Dawson}
\affiliation{Department of Astronomy \& Astrophysics, Center for Exoplanets and Habitable Worlds, The Pennsylvania State University, University Park, PA 16802, USA}

\author[0000-0001-6513-1659]{Jack J. Lissauer}
\affiliation{Space Science \& Astrobiology Division MS 245-3 NASA Ames Research Center Moffett Field, CA 94035, USA}

\author[0000-0003-1930-5683]{Man Hoi Lee}
\affil{Department of Earth Sciences, The University of Hong Kong, Pokfulam Road, Hong Kong}
\affil{Department of Physics, The University of Hong Kong, Pokfulman Road, Hong Kong}

\author[0000-0002-0436-7833]{Diana Kossakowski}
\affil{Max-Planck-Institut f\"{u}r Astronomie, K\"{o}nigstuhl  17, 69117 Heidelberg, Germany}

\author[0000-0003-3047-6272]{Felipe I.\ Rojas}
\affil{Instituto de Astrof\'isica, Facultad de F\'isica, Pontificia Universidad Cat\'olica de Chile, Chile}
\affil{Millennium Institute for Astrophysics, Chile}

\author[0000-0002-5945-7975]{Melissa J.\ Hobson}
\affil{Millennium Institute for Astrophysics, Chile}
\affil{Instituto de Astrof\'isica, Facultad de F\'isica, Pontificia Universidad Cat\'olica de Chile, Chile}

\author[0000-0001-8128-3126]{Paula Sarkis}
\affil{Max-Planck-Institut f\"{u}r Astronomie, K\"{o}nigstuhl  17, 69117 Heidelberg, Germany}

\author[0000-0001-8355-2107]{Martin Schlecker}
\affil{Max-Planck-Institut f\"{u}r Astronomie, K\"{o}nigstuhl  17, 69117 Heidelberg, Germany}

\author[0000-0002-8868-7649]{Bertram Bitsch}
\affil{Max-Planck-Institut f\"{u}r Astronomie, K\"{o}nigstuhl  17, 69117 Heidelberg, Germany}


\author[0000-0001-7204-6727]{Gaspar \'A. Bakos} 
\altaffiliation{Packard Fellow}
\affil{Department of Astrophysical Sciences, Princeton University, NJ 08544, USA}
\affil{Institute for Advanced Study, 1 Einstein drive, Princeton, NJ 08540, USA}


\author[0000-0001-8362-3462]{Mauro Barbieri} 
\affil{INCT, Universidad de Atacama, calle Copayapu 485, Copiap\'o, Atacama, Chile}

\author[0000-0002-0628-0088]{Waqas Bhatti} 
\affil{Department of Astrophysical Sciences, Princeton University, NJ 08544, USA}


\author{R. Paul Butler}  
\affiliation{Carnegie Institution for Science, Earth \& Planets Laboratory, 5241 Broad Branch Road NW, Washington DC 20015, USA}

\author[0000-0002-5226-787X]{Jeffrey D. Crane}  
\affiliation{The Observatories of the Carnegie Institution for Science, 813 Santa Barbara Street, Pasadena, CA 91101}


\author{Sangeetha Nandakumar} 
\affil{INCT, Universidad de Atacama, calle Copayapu 485, Copiap\'o, Atacama, Chile}

\author[0000-0002-2100-3257]{Mat\'ias R. D\'iaz} 
\affiliation{Departamento de Astronom\'ia, Universidad de Chile, Camino El Observatorio 1515, Las Condes, Santiago, Chile}
\affiliation{Las Campanas Observatory,Carnegie Institution for Science, Colina El Pino, Casilla 601, La Serena, Chile}

\author[0000-0002-8681-6136]{Stephen Shectman}
\affiliation{The Observatories of the Carnegie Institution for Science, 813 Santa Barbara Street, Pasadena, CA 91101}

\author{Johanna Teske} 
\affiliation{Carnegie Institution for Science, Earth \& Planets Laboratory, 5241 Broad Branch Road NW, Washington DC 20015, USA}


\author{Pascal Torres} 
\affil{Instituto de Astrof\'isica, Facultad de F\'isica, Pontificia Universidad Cat\'olica de Chile, Chile}
\affil{Millennium Institute for Astrophysics, Chile}

\author[0000-0001-7070-3842]{Vincent Suc} 
\affil{Instituto de Astrof{\'{i}}sica, Pontificia Universidad Cat{\'{o}}lica de Chile, Av. Vicu{\~{n}}a Mackenna 4860, 7820436 Macul, Santiago, Chile}

\author[0000-0002-1896-2377]{Jose I. Vines} 
\affiliation{Departamento de Astronom\'ia, Universidad de Chile, Camino El Observatorio 1515, Las Condes, Santiago, Chile}

\author[0000-0002-6937-9034]{Sharon X.~Wang} 
\affiliation{The Observatories of the Carnegie Institution for Science, 813 Santa Barbara Street, Pasadena, CA 91101}
\affiliation{Department of Astronomy, Tsinghua University, Beijing 100084, People's Republic of China}

\author[0000-0003-2058-6662]{George R. Ricker} 
\affiliation{Department of Physics and Kavli Institute for Astrophysics and Space Research, Massachusetts Institute of Technology, Cambridge, MA 02139, USA}

\author[0000-0002-1836-3120]{Avi Shporer} 
\affiliation{Department of Physics and Kavli Institute for Astrophysics and Space Research, Massachusetts Institute of Technology, Cambridge, MA 02139, USA}

\author{Andrew Vanderburg} 
\affil{Department of Astronomy, University of Wisconsin-Madison, Madison, WI 53706, USA}

\author{Diana Dragomir} 
\affiliation{Department of Physics and Astronomy, University of New Mexico, 1919 Lomas Blvd NE, Albuquerque, NM 87131, USA}

\author{Roland Vanderspek} 
\affiliation{Department of Physics and Kavli Institute for Astrophysics
and Space Research, Massachusetts Institute of Technology, Cambridge, MA
02139, USA}

\author[0000-0002-7754-9486]{Christopher~J.~Burke}
\affiliation{Department of Physics and Kavli Institute for Astrophysics and Space Research, Massachusetts Institute of Technology, Cambridge, MA 02139, USA}

\author[0000-0002-6939-9211]{Tansu~Daylan}
\altaffiliation{Kavli Fellow}
\affiliation{Department of Physics and Kavli Institute for Astrophysics and Space Research, Massachusetts Institute of Technology, Cambridge, MA 02139, USA}

\author{Bernie~Shiao}
\affiliation{Space Telescope Science Institute, 3700 San Martin Drive, Baltimore, MD, 21218, USA}

\author[0000-0002-4715-9460]{Jon M. Jenkins} 
\affiliation{NASA Ames Research Center, Moffett Field, CA 94035, USA}

\author[0000-0002-5402-9613]{Bill Wohler} 
\affiliation{NASA Ames Research Center, Moffett Field, CA 94035, USA}
\affiliation{SETI Institute, Mountain View, CA 94043, USA}

%

\author[0000-0002-6892-6948]{Sara Seager} 
\affiliation{Department of Physics and Kavli Institute for Astrophysics
and Space Research, Massachusetts Institute of Technology, Cambridge, MA
02139, USA}
\affiliation{Department of Earth, Atmospheric and Planetary Sciences,
Massachusetts Institute of Technology, Cambridge, MA 02139, USA}
\affiliation{Department of Aeronautics and Astronautics, MIT, 77
Massachusetts Avenue, Cambridge, MA 02139, USA}

\author[0000-0002-4265-047X]{Joshua N. Winn} 
\affiliation{Department of Astrophysical Sciences, Princeton University,
NJ 08544, USA}

 
\begin{abstract}
TOI-2202\,b is a transiting warm Jovian-mass planet with an orbital period of P=11.91 days identified from the Full Frame Images data of five different sectors of the $\tess$ mission.  Ten $\tess$ transits of TOI-2202\,b combined with three follow-up light curves obtained with the CHAT robotic telescope show strong transit timing variations (TTVs) with an amplitude of about 1.2 hours. Radial velocity follow-up with FEROS, HARPS and PFS confirms the planetary nature of the transiting candidate (a$_{\rm b}$ =  0.096  $\pm$ 0.002 au,  
m$_{\rm b}$ = 0.98 $\pm$ 0.06 M$_{\rm Jup}$), and dynamical analysis of RVs, transit data, and TTVs points to an outer Saturn-mass companion (a$_{\rm c}$ =  0.155  $\pm$ 0.003 au, m$_{\rm c}$= $0.37 \pm 0.10$ M$_{\rm Jup}$) near the 2:1  mean motion resonance. Our stellar modeling indicates that TOI-2202 is an early K-type star with a mass of 0.82 M$_\odot$, a radius of 0.79 R$_\odot$, and solar-like metallicity. The TOI-2202 system is very interesting because of the 
two warm Jovian-mass planets near the 2:1 MMR, which is a rare configuration, and their formation and dynamical evolution are still not well understood.  
\end{abstract}

\keywords{Techniques: radial velocities $-$ Planets and satellites: detection, dynamical evolution and stability 
   $-$ (Stars:) planetary systems
}

\section{Introduction}
\label{sec1}

The past twenty-five years of exoplanet searches have resulted in over 4300 confirmed planets, including over 700\footnote{up to date list
available on \url{https:\\exoplanet.eu}} systems with multiple planets. 
The physical characteristics of the discovered exoplanet systems 
show a great contrast with the Solar System.
Of course, the observed diversity of exoplanet populations is still a subject of observational biases \citep[see e.g.,][]{Fischer2014}. 
For instance, the two most successful exoplanet detection techniques --- 
the transit method and the radial velocity (RV) method --- 
are capable of detecting short period planets as small as Earth. 
The longest-running RV surveys have a sufficient temporal baseline to detect long-period planets \citep{Bonfils2013, Reffert2015, Butler2017,  Udry2019, Wittenmyer2020}, but the achievable precision is only sufficient to detect Jovian planets, or at best, Saturn-mass planets.
But despite the known biases, we can still use the current observational exoplanet data to fine-tune the applicable planet formation theories in an attempt to understand 
the planet formation mechanisms in~general.

The discovery of very close-orbiting planets, planets on eccentric orbits, and pairs of planets in mean-motion resonances (MMR) have led to major developments in the theory of the formation and dynamical evolution of planets, and
in particular, in our understanding of the importance of interactions between planets and the protoplanetary disk \citep[][]{Ida2010, Kley2012,Coleman2014, Baruteau2014,Levison2015,Bitsch2020, Schlecker2020b,Matsumura2021}.
One of the long-standing challenges in the field of exoplanets is to explain the origin of the population of giant planets with orbits interior to the so-called snow line. These objects are not easily understood within standard formation models that require rapid accretion of gas by a solid embryo before the stellar radiation dissipates the gas from the protoplanetary disc. This rapid, solid accretion is favored beyond the snow line. Giant planets are expected then to migrate from a couple of astronomical units to the inner regions of the system to produce the population of hot ($P<$ 10\,d) and warm (10\,d $<P<$ 300\,d) Jovian mass planets. Typical migration mechanisms can be divided in two groups, namely: disc migration \citep[e.g.,][]{Lin1986}, and high eccentricity tidal migration \citep[e.g.,][]{Rasio1996,Fabrycky2007,Bitsch2020}. Both types of mechanisms predict significantly different orbital configurations for the migrating planet, and the characterization of these properties, particularly on warm Jovians \citep[][]{Huang2016, Petrovich2016,Santerne2016,Dong2021}, can be used to constrain migration theories.

Gas accretion is thought to be faster beyond the ice line because the cores would be large enough. Nonetheless, given the high-frequency of sufficiently large "cores" (i.e., super-Earths and mini-Neptunes) discovered close to their star, the in-situ formation of giant planets is also a possible scenario \citep[e.g.,][]{Batygin2016}. In addition, planets can also be scattered towards warmer orbits via violent  gravitational interactions and instabilities \citep[e.g.,][]{Ford2005}.

In this context, it is fundamentally important to measure the dynamical mass and orbital eccentricity of the warm Jovian planets. For many systems, this can only be achieved by combining precise transit and RV observational exoplanet data. 
NASA's {\it Transiting Exoplanet Survey Satellite} \citep[$\tess$;][]{Ricker2015}
aims to detect planets through the transit method around relatively bright stars that are suitable for precise Doppler follow-up to determine the planetary mass, radius, and bulk density, among other important physical parameters. $\tess$ has already led to more than 130 new discovered planets, most of which were confirmed by Doppler spectroscopy \citep[e.g.,][among many]{Wang2019, Trifonov2019a, Dumusque2019, Luque2019,  kossakowski:2019, Teske2020, Schlecker2020, Espinoza2020}.

 \begin{figure*}[tp]
    \centering

    \includegraphics[width=5.9cm]{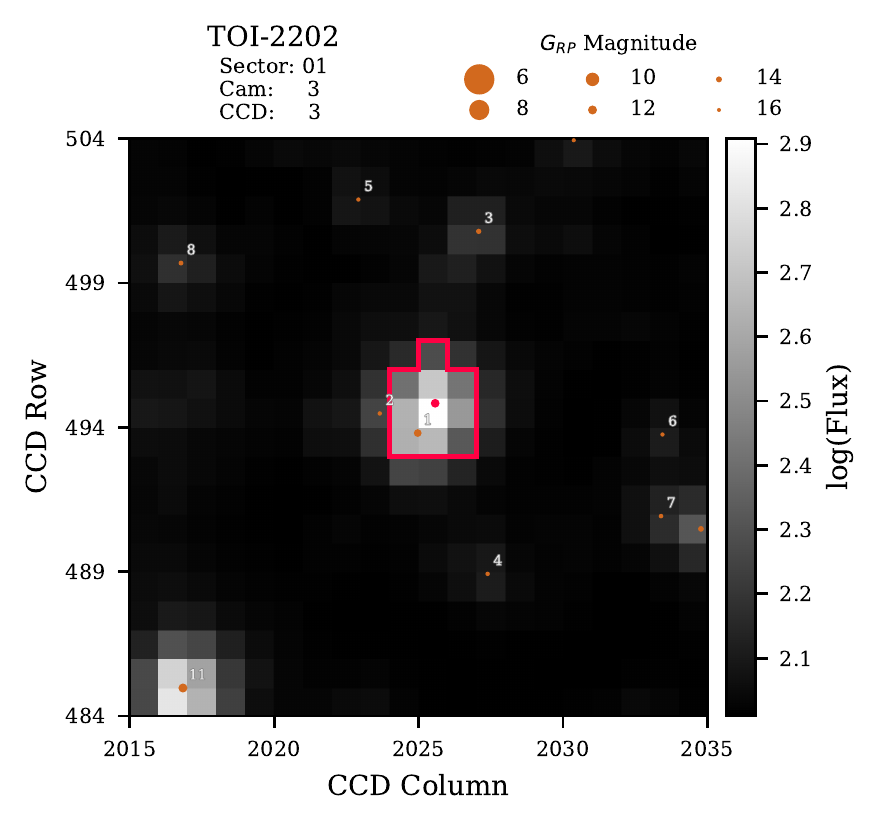} 
    \includegraphics[width=5.9cm]{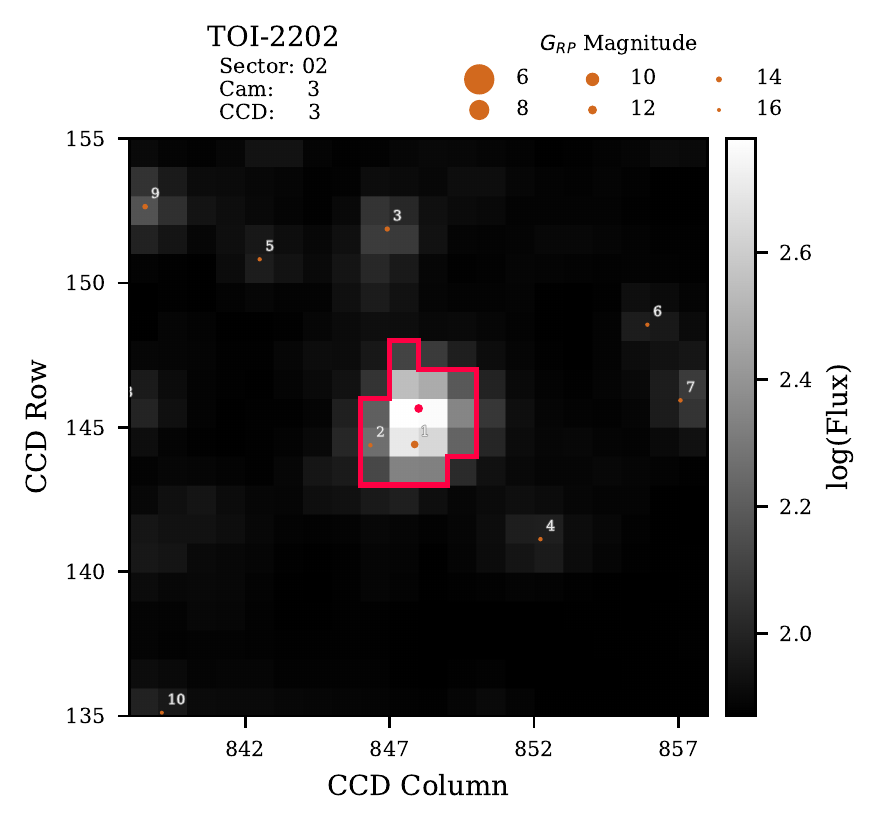} 
    \includegraphics[width=5.9cm]{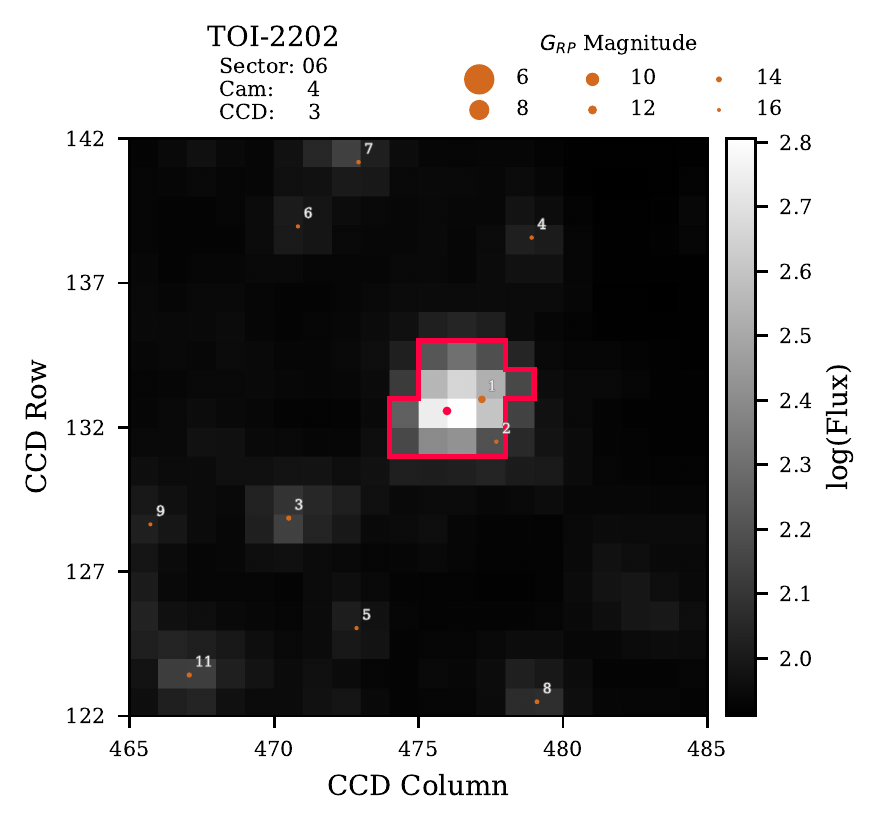} \\ 
    \includegraphics[width=5.9cm]{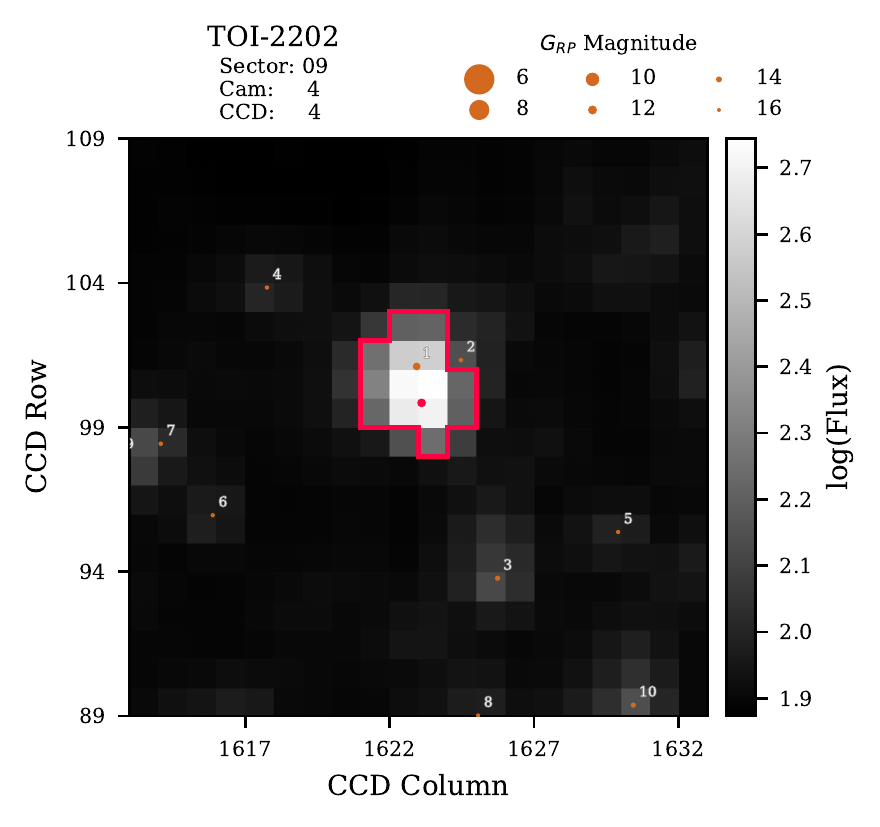} 
    \includegraphics[width=5.9cm]{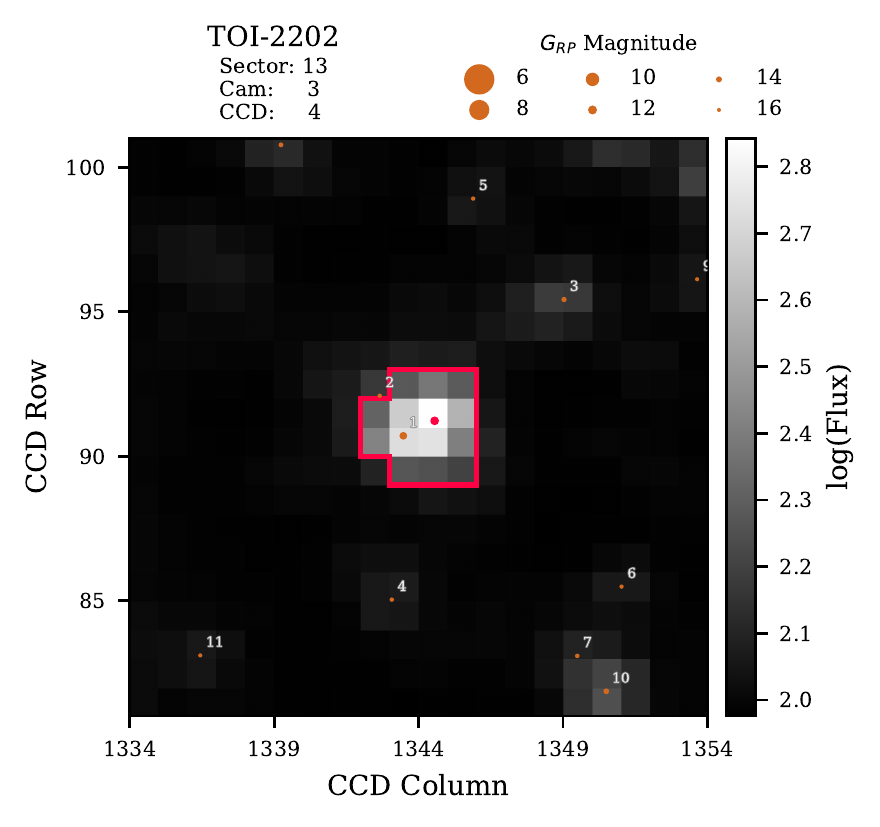}

    \caption{Target pixel file (TPF) image of TOI-2202 in $\tess$ Sector 2. The red dots show the position of TOI-2202 (brighter) and the neighbor star TIC\,358107518 (fainter). The red borders in the pixel space are the one used in to construct the 
    $\tess$ Simple Aperture Photometry (SAP). {\em Gaia} targets are marked with orange circles, size coded by their G magnitude.
    }
    \label{RV_results} 
\end{figure*}

In this paper, we report the discovery of a warm Jovian-mass planet pair around the K-dwarf star TOI-2202 (TIC\,358107516\footnote{The target became a $\tess$ Object of Interest \citep[TOI,][]{Guerrero2021}, while this work was in an advanced stage. Our discovery is based on the $\tess$ Full Frame Image data adopting the target designation TIC\,358107516 (see \autoref{Sec3.1.1}). Consequently, we adopted the TOI-2202 designation for consistency with the $\tess$ survey.}).
The inner planet TOI-2202\,b shows unambiguous 
transit events with a period of 11.9 days, recorded by $\tess$ and ground-based photometry. 
The strong transit timing variations (TTVs) of the transiting planet and the precise radial velocity measurements we obtained for this target revealed the existence of an additional outer Saturn-mass planet TOI-2202\,c with an orbital period 
of 24.7 days, thus forming a planet pair close to the 2:1 MMR.  
 This discovery was made in the context of the Warm gIaNts with tEss (WINE) collaboration, which focuses on the systematic characterization of $\tess$ transiting warm giant planets \citep[e.g.,][]{hd1397,jordan:2020,brahm:2020,Schlecker2020}.

In \autoref{sec2} we present our stellar parameter estimates of TOI-2202.
In \autoref{sec3} we present the transit photometry and Doppler 
observational data used to characterize the TOI-2202 multiple planet system.
In \autoref{sec4} we describe our orbital analysis using a self-consistent dynamical modeling scheme, whereas in \autoref{sec5} we comment on the dynamical and long-term stability properties of the TOI-2202 system.
\autoref{sec6} is for our summary and conclusions.

\section{Data}
\label{sec2}

\begin{figure*}[tp]
    \centering
    \includegraphics[width=18cm]{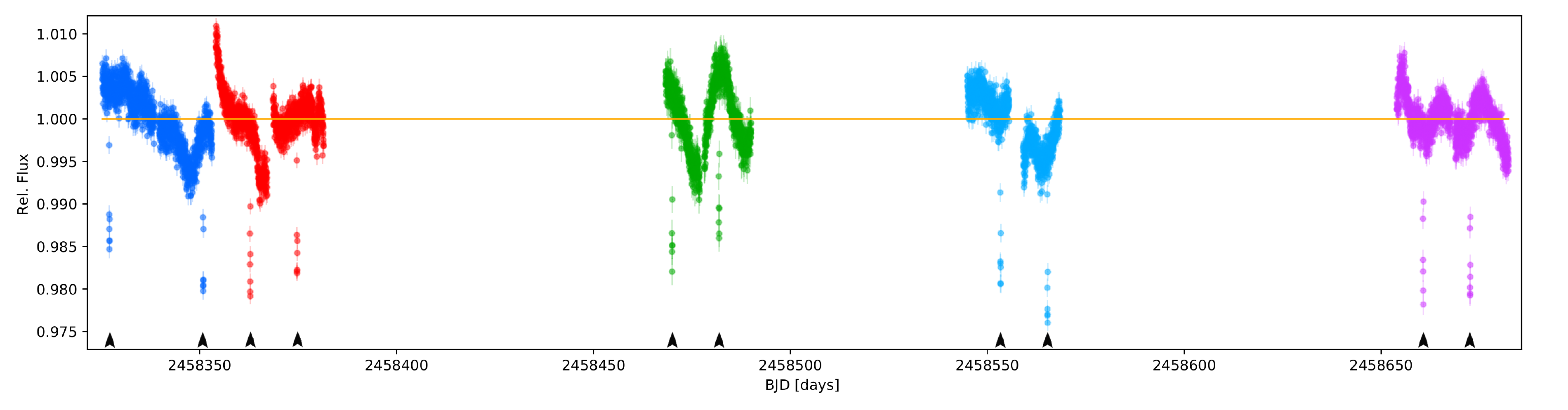} \\
    \includegraphics[width=18cm]{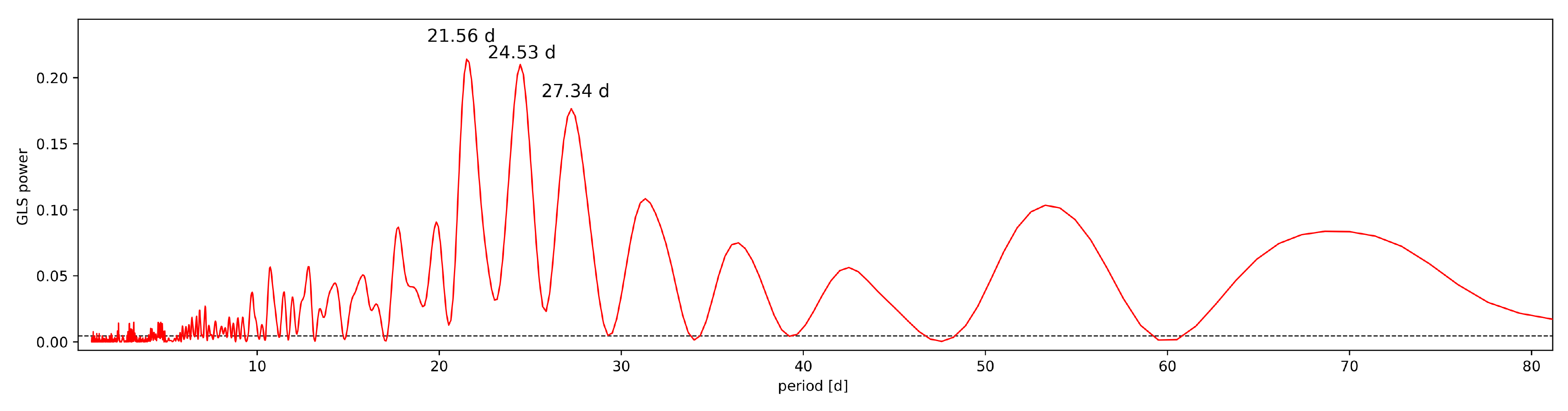}  \\ 
\caption{The top panel shows the raw $\tess$ photometry data of TOI-2202 reduced with {\tt tesseract} and normalized to its median. $\tess$ data from Sectors: 1 (blue), 2 (red), 6 (green), 9 (cyan), and 13 (magenta) clearly show transit events with a period of $\sim$ 11.91\,days, but also exhibit periodic systematics, which could be attributed to stellar activity, instrumental effects or combination of both. The bottom panel shows the GLS periodogram of the raw $\tess$ data, yielding  prominent peaks in the range between 20-30\,days. The dashed line is the 0.1\% FAP. 
}
 
\label{fig2} 
\end{figure*}

\subsection{Transit photometry}
\label{Sec3.1}

\subsubsection{TESS}
\label{Sec3.1.1}

TOI-2202 was observed in five out of the 13 Sectors of the first year of the $\tess$ primary mission. Observations were performed with the 30\,minute
cadence mode in Sectors 1, 2, 6, 9, and 13 between July 2018 and July 2019.
TOI-2202 b was identified in the light curves extracted from the $\tess$ FFIs.
This was done using a pipeline called \texttt{tesseract}\footnote{\url{https://github.com/astrofelipe/tesseract}} (Rojas et al. in prep). 
\texttt{tesseract} receives the TIC ID as input and performs simple aperture photometry on the FFIs via the \texttt{TESSCut}~\citep{TESSCut} and \texttt{lightkurve} \citep{lightkurve} packages. In the context of the WINE collaboration, we generated light curves for all Sectors of the first year of the $\tess$ primary mission for stars of the TICv8 catalog brighter than T = 13 mag. Transiting candidates are identified by running 
the \texttt{transitleastsquares} package \cite[TLS;][]{Hippke2019b} on each light curve, and also by searching for individual negative deviations from the median flux. This latter processing allowed us to identify long period planets \citep[e.g.,][]{Schlecker2020} and single transiters \citep{gill:2020}. TOI-2202 was identified as a candidate in individual $\tess$ light curves using both detection methods, finding a periodicity of the transits of $\approx 12$ days.
\autoref{RV_results} shows the target pixel file (TPF) image of TOI-2202\,constructed from the $\tess$ image frames and {\em Gaia} DR2 data. We investigated if the transit signal was coming from neighboring stars by generating light curves for each of the pixels in a region 20 pixels wide around the target star, and also by analyzing the time series associated to the background flux. We found that the signal indeed originated close ($<2'$) to the target star. Nonetheless, as is show in \autoref{RV_results}, TOI-2202 has a slightly fainter (T=12.7) neighboring star located at $\approx 24\arcsec$ from it (TIC358107518) that was not possible to fully reject as the source of the observed transits. Our ground based photometric and Doppler follow up confirmed that the source of the transit signal was indeed TOI-2202 (see \autoref{CHAT} and \autoref{RVdata}).

Since {\tt tesseract} does not correct for contamination in the TESS apertures from nearby stars, we also 
apply dilution correction for the contamination of TIC\,358107518. To estimate the dilution, we use the $R-P$ Gaia DR3 fluxes of TOI-2202 and TIC\,358107518, and equation (6) in \citet{Espinoza2019}. For TOI-2202 the mean $R-P$ flux is 113856 electron\,s$^{-1}$, while the mean $R-P$ flux of TIC\,358107518 is 66547 electron\,s$^{-1}$,, and we obtain a dilution factor of 0.63, which was applied to the light curves. 

The top panel of \autoref{fig2} shows the dilution-corrected, median-normalized, $\tess$ FFI light curves of TOI-2202.
All the $\tess$ 30-minute
cadence FFI data taken in Sectors 1, 2, 6, 9, and 13 suffered from notable systematics, which could be attributed to stellar activity, instrumental effects, or a combination of both.
The transit events with a depth of approximately 1\%, which was consistent with being produced by a Jupiter-sized planet with a period of about 11.9\,d are can be also easily identified on the combined $\tess$ light curve.
The bottom panel of \autoref{fig2} shows a GLS periodogram 
of the $\tess$ light curve yielding prominent peaks in the range between 20-30\,days, which come from the light curve systematics, and which 
are generally in line with the most-likely rotational period range of TOI-2202 (see \autoref{sec3}).


\subsubsection{CHAT}
\label{CHAT}
After we detected the transit events in the $\tess$ FFIs of TOI-2202, we scheduled a photometric monitoring from the ground with the 0.7m Chilean-Hungarian Automated Telescope (CHAT) telescope installed in Las Campanas Observatory, in Chile. We obtained four light curves on four different nights between February and November 2019. All observations were performed in the Sloan $i'$ band with exposure times of 150s. The data were reduced and processed into differential photometry light curves with a dedicated package \citep[e.g.,][]{kossakowski:2019, jordan:2019}. 

The transit signal of the TOI-2202\,b planet candidate, was detected on three of the four light curves. (two partial transits in ingress, and a single full transit), which allowed us to confirm that the transits do not occur on the slightly fainter close companion, to refine the transit parameters, and to further characterize the transit timing variations of the system. The CHAT follow-up data are available in electronic form in 
ExoFOP, at \url{https://exofop.ipac.caltech.edu/tess/target.php?id=358107516}, and the light curves are displayed in \autoref{tra_data}.


\subsection{RV data}
\label{RVdata}

\subsubsection{FEROS}

We conducted a spectroscopic follow-up campaign for TOI-2202 between February and November 2019 with the FEROS spectrograph \citep{Kaufer1999} installed at the ESO-MPG 2.2m telescope in La Silla Observatory.
Our immediate objective of the FEROS observations was to determine if the transit-like signals
present on the stellar light curve were indeed produced by a transiting Jovian companion. In total we obtained 26 FEROS spectra with exposure times of 1200 sec and 1500 sec, delivering signal-to-noise (S/N) ratios per resolution element ranging from 20 to 60. Observations were performed with the simultaneous calibration technique, in conjunction with a ThAr calibration lamp.
FEROS data were reduced, extracted and analyzed with the \texttt{ceres} pipeline \citep{ceres} delivering RV and bisector span measurements with a typical uncertainty of $\sim$15 m\,s$^{-1}$. The obtained RVs are tabulated in \autoref{table:TIC358107516_FEROS}, and displayed in \autoref{TTV_plot1} and \autoref{TTV_plot2} as a function of time and orbital phase. These FEROS RVs allowed the identification of a Keplerian signal with a period consistent with that of the transiting events of $\tess$ and CHAT, and an amplitude compatible with a Jovian mass object. 

To further check that the close neighbor companion star TIC358107518 was not the source of the transit signal, we also obtained three FEROS spectra of this star in March 2019, taken at phases 0.2, 0.5, and 0.8. We find no significant RV variations for this object, and thus we concluded that TOI-2202 is indeed the source of the transiting signal in the $\tess$ light curve.

\begin{table}[tp]

\caption{Stellar parameters of TOI-2202 and their 1$\sigma$ uncertainties derived using 
ZASPE spectral analyses, {\em Gaia} parallax, broad band photometry and PARSEC models. 
The values in parentheses are floor uncertainties predicted by \citet{Tayar2020}
and adopted in our work.
}
\label{table:phys_param}    


\centering          
\begin{tabular}{ p{3.0cm} l r}     
\hline\hline  \noalign{\vskip 0.5mm}        
  Parameter   & TOI-2202   &  reference \\  
\hline    \noalign{\vskip 0.5mm}                   
   Spectral type                            & K8V          & [1] \\ 
   Distance  [pc]                           & 235.93$_{-1.04}^{+1.05}$   & [2,3] \\   
   Mass    [$M_{\odot}$]                    & 0.823$_{-0.023}^{+0.027}$ (0.041)   & This paper\\
   Radius    [$R_{\odot}$]                  & 0.794$_{-0.007}^{+0.007}$ (0.032)   & This paper  \\
   Luminosity    [$L{_\odot}$]              & 0.397$_{-0.013}^{+0.014}$  (0.014)   & This paper \\
   Age    $[$Gyr$]$                         & 7.48$_{-3.33}^{+3.32}$     &  This paper\\  
   A$_V$   [mag]       & 0.242$_{-0.054}^{+0.0.056}$ & This paper     \\
   $T_{\mathrm{eff}}$~[K]                   & 5144 $\pm$ 50 (103)   & This paper \\
   $\log g~[\mathrm{cm\cdot s}^{-2}]$       & 4.55 $\pm$ 0.20    & This paper \\   
   {}[Fe/H]                                 & 0.04 $\pm$ 0.05   & This paper  \\
   $v\cdot\sin(i)$~$[$km\,s$^{-1}]$              & 1.7  $\pm$ 0.5    & This paper   \\                                          
    
\hline\hline \noalign{\vskip 0.5mm}   

\end{tabular}
 


\tablecomments{\small  
[1] \citet{ESA}, 
[2] \citet{Gaia_Collaboration2016, Gaia_Collaboration2018b} , 
[3] \citet{Bailer_Jones}. 
 }

\end{table}


\subsubsection{HARPS}

The High Accuracy Radial velocity Planet Searcher \citep[HARPS,][]{Mayor2003} is
an ultra-stable high-resolution ($R$=115,000) echelle spectrograph 
mounted at the 3.6\,m telescope of the European Southern Observatory (ESO) in La Silla, Chile.  HARPS is capable of delivering stellar RV measurements with a precision down to $\sim$1\,m\,s$^{-1}$.

We obtained 21 spectra with HARPS between June 2019 and February 2020. We adopted an exposure time of 1800 sec, which translated in spectra with a S/N of $\sim$ 25. We retrieved precise RV measurements derived by ESO-DRS pipeline, which uses a spectrum cross-correlation function (CCF) method with a weighted binary mask \citep{Pepe2002}. The DRS pipeline also provides CCF's full-width half-maximum (FWHM) and the Bisector Inverse Slope span (BIS-span) measurements, which are valuable stellar activity indicators \citep[][]{Queloz2001}. 
Additionally, we derived precise RVs and stellar activity indicators 
from the HARPS spectra of TOI-2202 with the 
SpEctrum Radial Velocity AnaLyser \citep[SERVAL,][]{Zechmeister2018} pipeline.
SERVAL also measures stellar activity indicators such as H$\alpha$,
Na\,I D, Na\,II D and the differential line width (dLW), quantifying variations in the spectral line widths, and the chromatic RV index (CRX) of the spectra \citep[for detail description of 
the SERVAL activity time series measurements see][]{Zechmeister2018}.
We find that in the case of TOI-2202, the original RVs derived with the ESO-DRS pipeline are somewhat less precise, but overall, more accurate. For instance, the median RV uncertainty of ESO-DRS is $\hat{\sigma_{\rm DRS}}$ = 7.09 m\,s$^{-1}$ and for SERVAL the median RV uncertainty is only $\hat{\sigma_{\rm SERVAL}}$ = 3.26 m\,s$^{-1}$, but the latter dataset contains three very strong outliers at epochs BJDs = 2458773.7777, 2458774.7145, and 2458838.6934. While the rest of the SERVAL and the DRS RVs are generally consistent, we decide to use the larger, more consistent, DRS RV data set for the orbital analysis in our study.
The precise RVs and activity index data from SERVAL and DRS are tabulated in \autoref{table:TIC358107516_HARPS}, and will be also available in electronic form in the {\tt HARPS-RVBank ver.02}\footnote{\url{https://github.com/3fon3fonov/HARPS_RVBank}, \\ \url{https://www2.mpia-hd.mpg.de/homes/trifonov/HARPS_RVBank.html} }
\citep{Trifonov2020}. 
A visual inspection of the HARPS-DRS is shown in \autoref{TTV_plot1} and \autoref{TTV_plot2} as a function of time and orbital phase.

\subsubsection{PFS}

TOI-2202 was also monitored with the Planet Finder Spectrograph \citep{crane2006,crane2008,crane2010} installed at the 6.5m Magellan/Clay telescope at Las Campanas Observatory. TOI-2202 was observed with the iodine cell in four different epochs between July and December of 2019, adopting an exposure time of 1200 sec, and using 3$\times$3 CCD binning mode to minimize read-noise. TOI-2202 was also observed without the iodine cell in order to generate the template for computing the radial velocities, which were derived following the methodology of \cite{butler1996}. The PFS radial velocities are presented in \autoref{table:TIC358107516_PFS}  and displayed in \autoref{TTV_plot1} and \autoref{TTV_plot2}.

\section{Stellar parameters of TOI-2202}
\label{sec3}

The atmospheric parameters of TOI-2202 were computed from the co-added HARPS spectra using the \texttt{zaspe} package \citep{zaspe}, which delivers $T_{\rm{eff}}$, $\log{g}$, [Fe/H], and $v\sin{i}$ through comparison against a grid of synthetic spectra generated from the ATLAS9 model atmospheres \citep{atlas9}.

The physical parameters of TOI-2202 were obtained by using the PARSEC evolutionary models, following the prescription described in \citet{hd1397}. These models allow us to compare the absolute magnitudes for a given set of stellar parameters to those of the target star by using the distance to the star as obtained from the {\em Gaia} DR2 catalog \citep{Gaia_Collaboration2018b}. For this comparison we used the {\em Gaia} G, G$_{BP}$, and G$_{RP}$; and the 2MASS, $J$, $H$, and $K_{s}$ bands.  We fixed the metallicity to the value found with \texttt{zaspe}, and explore the parameter space for the stellar age and M$_{\star}$ by using the {\tt emcee} package \citep{emcee}.
The results of this analysis allow us to determine a more precise value for the $\log{g}$ than that of the spectroscopic analysis. We therefore apply an iterative process involving these two procedures in which the output $\log{g}$ of the SED analysis is given as input for a new \texttt{zaspe} run, until reaching convergence.

 \begin{figure}[tp]
    \centering
    \includegraphics[width=9cm]{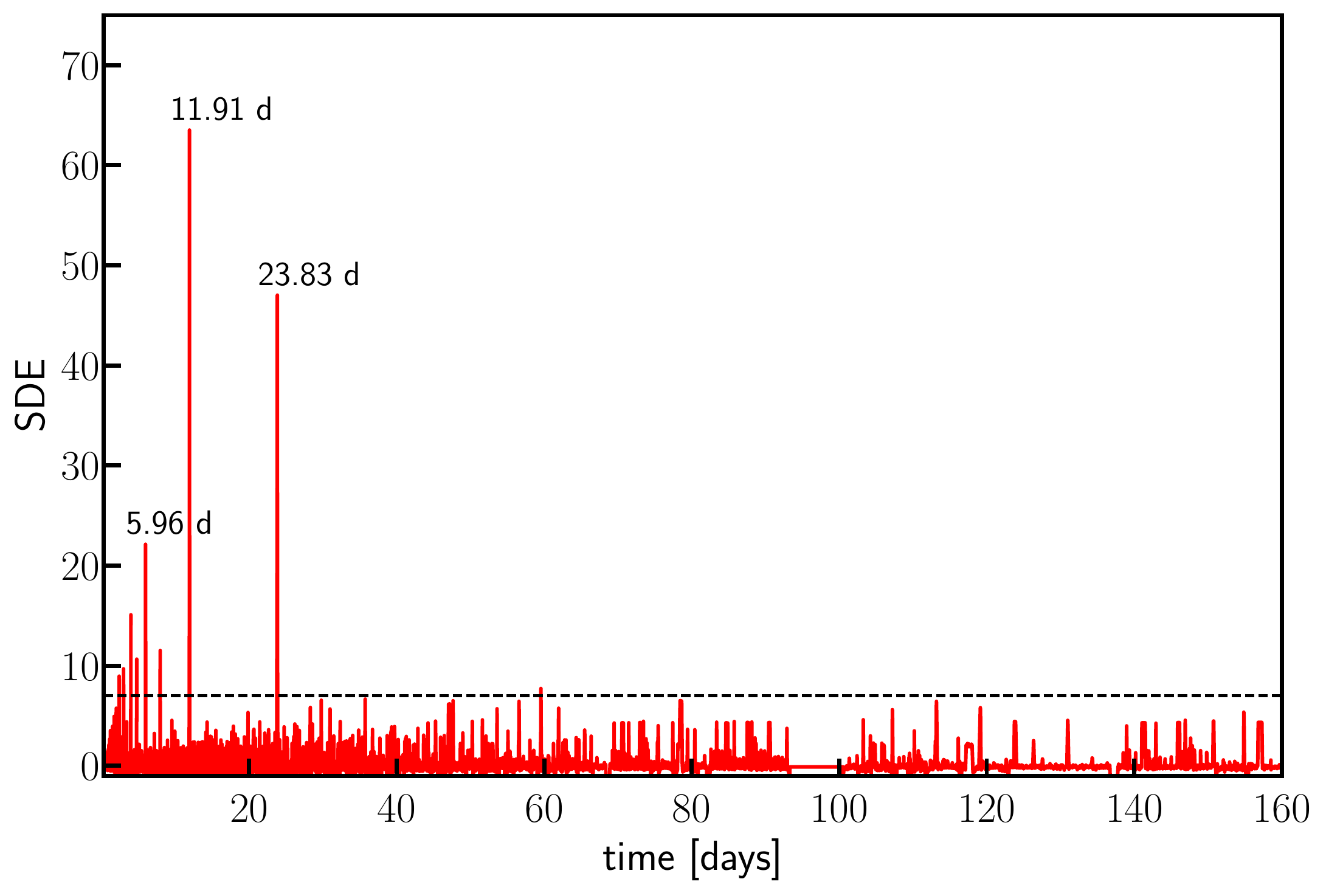} 
 
    \caption{
    TLS power spectra of the detrended FFI $\tess$ and CHAT light curve data of TOI-2202.    
    The planetary transit signal at $P_b=11.91$\,d is accompanied by harmonics at 5.96\,d and 23.83\,d, etc.
    Horizontal dashed line indicates the signal detection efficiency (SDE) power level of 7.0, which corresponds to the TLS false positive rate of 1\%.
    }
    \label{TLS_results} 
\end{figure}

Our \texttt{zaspe} analysis shows that TOI-2202 is probably a late K-type dwarf star with a mass of 0.823$_{-0.023}^{+0.027}$ $M_{\odot}$,
radius of 0.794$_{-0.007}^{+0.007}$ $R_{\odot}$, and solar-like metallicity. The full set of atmospheric and physical parameters are listed in \autoref{table:phys_param}. However, we note that our relatively small uncertainties in the stellar parameters are internal and do not include possible systematic differences with respect to other stellar models. 
Therefore, we inflate the stellar uncertainties to more realistic values 
following the prescription of \citet{Tayar2020}. These authors found that 
for a main-sequence star like TOI-2202, the publicly available model grids suggest systematic uncertainty floor of order $\sim$5\% in mass, $\sim$4\% in radius and $\sim$2\% in temperature and luminosity, respectively \citep[see,][for more details]{Tayar2020}. The inflated uncertainty estimates are also listed in \autoref{table:phys_param}, and are the adopted stellar parameter uncertainties through this work. 

With an estimated radius of 0.794$\pm$0.032 $R_{\odot}$ and $v\cdot\sin(i)$ 1.7  $\pm$ 0.5 [$km\,s^{-1}$], the most likely stellar rotation period is 26.8$\pm$9.0 days, which agrees well with the observed systematic TESS light curve periodicity, which is likely induced by stellar activity (see \autoref{fig2}).

\section{Analysis and Results}
\label{sec4}

\subsection{Tools}
\label{Sec4.1}

For data and orbital analysis, we employed the {\tt Exo-Striker} exoplanet toolbox\footnote{\url{https://github.com/3fon3fonov/exostriker}}
\citep{Trifonov2019_es}.
The {\tt Exo-Striker} provides easy access to a large variety of public tools for exoplanet data analysis, such as: 
a generalized Lomb-Scargle periodogram \citep[GLS;][]{Zechmeister2009},
a maximum likelihood periodogram \citep[MLP;][]{Baluev2008, Baluev2009, Zechmeister2019},
transit photometry de-trending via {\tt wotan} \cite[][]{Hippke2019}, 
and transit period search 
via the {\tt transitleastsquares} package \cite[TLS;][]{Hippke2019b}, 
which we used in this work for RV and transit photometry signal analysis. 
The {\tt Exo-Striker} is able to model data with a standard Keplerian model or with a more complex dynamical model in case of gravitationally interacting multiple-planet systems detected on RVs or transit photometry data. 
The {\tt Exo-Striker} works in Jacobi coordinate system, which is a natural frame for orbital parameterization of multiple-planet systems \citep[][]{Lee2003}.
The modeling can be performed either by `best-fit' optimization schemes (i.e. Levenberg-Maquardt, Nelder-Mead, Newton, etc.), 
or sampling schemes such as an affine-invariant ensemble Markov Chain Monte Carlo (MCMC) sampler \citep{Goodman2010} 
via the \texttt{emcee} package \citep{emcee}, and the nested sampling technique \citep{Skilling2004} via {\tt dynesty} sampler \citep{Speagle2020}.

\begin{table}[ht]

\centering   
\caption{{Individual mid-transit time estimates of TOI-2202\,b extracted from $\tess$ and CHAT used for TTV analysis.}} 
\label{table:TTVdata}

\begin{tabular}{ccccrrrr}     

\hline\hline  \noalign{\vskip 0.7mm}
N Transit \hspace{0.0 mm}& t$_{\rm 0}$ [BJD] & $\sigma$ t$_{\rm 0}$ [BJD] & Instrument \\
\hline \noalign{\vskip 0.7mm}

1  & 2458327.102860 & 0.001263  & $\tess$ \\
3  & 2458350.929974 & 0.000904  & $\tess$ \\
4  & 2458362.843557 & 0.000851  & $\tess$ \\
5  & 2458374.749429 & 0.001078  & $\tess$ \\
13 & 2458469.975553 & 0.001690  & $\tess$ \\
14 & 2458481.884441 & 0.001284  & $\tess$ \\
20 & 2458553.368053 & 0.000957  & $\tess$ \\
21 & 2458565.287934 & 0.001310  & $\tess$ \\
29 & 2458660.678263 & 0.001081  & $\tess$ \\
30 & 2458672.594730 & 0.000921  & $\tess$ \\
37 & 2458755.954035 & 0.000926  & CHAT \\
38 & 2458767.856769 & 0.001243  & CHAT \\
40 & 2458791.661596 & 0.000718  & CHAT \\

\hline \noalign{\vskip 0.7mm}

\end{tabular}

\end{table}

The RV modeling schemes are intrinsic\footnote{ Some of the available RV routines were originally developed by \citet{Tan2013} for the analysis of the 2:1 MMR system HD\,82943, and were further developed for the analysis of the $\eta$\,Ceti, HD\,59686, and HD\,202696 systems \citep{Trifonov2014,Trifonov2018b,Trifonov2019b} } to the {\tt Exo-Striker}, while currently,
for transit light curve models, the {\tt Exo-Striker} employs the BAsic Transit Model cAlculatioN package \citep[{\tt batman}; ][]{Kreidberg2015}. 
In addition, dynamical modeling of TTV data is done with a Python wrapper of the TTV-fast package \citep{Deck2014}. 
When needed, RV and transit data can be 
additionally modeled, with Gaussian process (GP) regression models using the {\tt celerite} package \citep{Mackey2017}, which is also included in the {\tt Exo-Striker}.

\subsection{Transit light curve analysis}
\label{Sec4.2}

We inspected the $\tess$ FFI light curves of TOI-2202 derived with {\tt tesseract}. 
For our preliminary transit light curve analysis, we further de-trend each of the {\tt tesseract} sector light curves 
with a robust (iterative) Mat\'ern GP kernel, which was aimed to capture the 
systematic variation of the light curves \citep[see][]{Hippke2019}. 
For CHAT data, we applied additional transit photometry de-trending as a function of the airmass at the epoch of observations.
Our de-trending scheme resulted in nearly flat, normalized, $\tess$ and CHAT light curves, which we inspect for transits using the 
TLS algorithm. \autoref{TLS_results} shows the constructed TLS power-spectra of the available transit data of TOI-2202.    
A significant peak occurs at a period of $P_b=11.91$\,d, followed by peaks at 5.96\,d and 23.83\,d, etc., which are simply low-order
harmonics of the actual transit signal.

 \begin{figure}[tp]
    \centering
    \includegraphics[width=8.5cm]{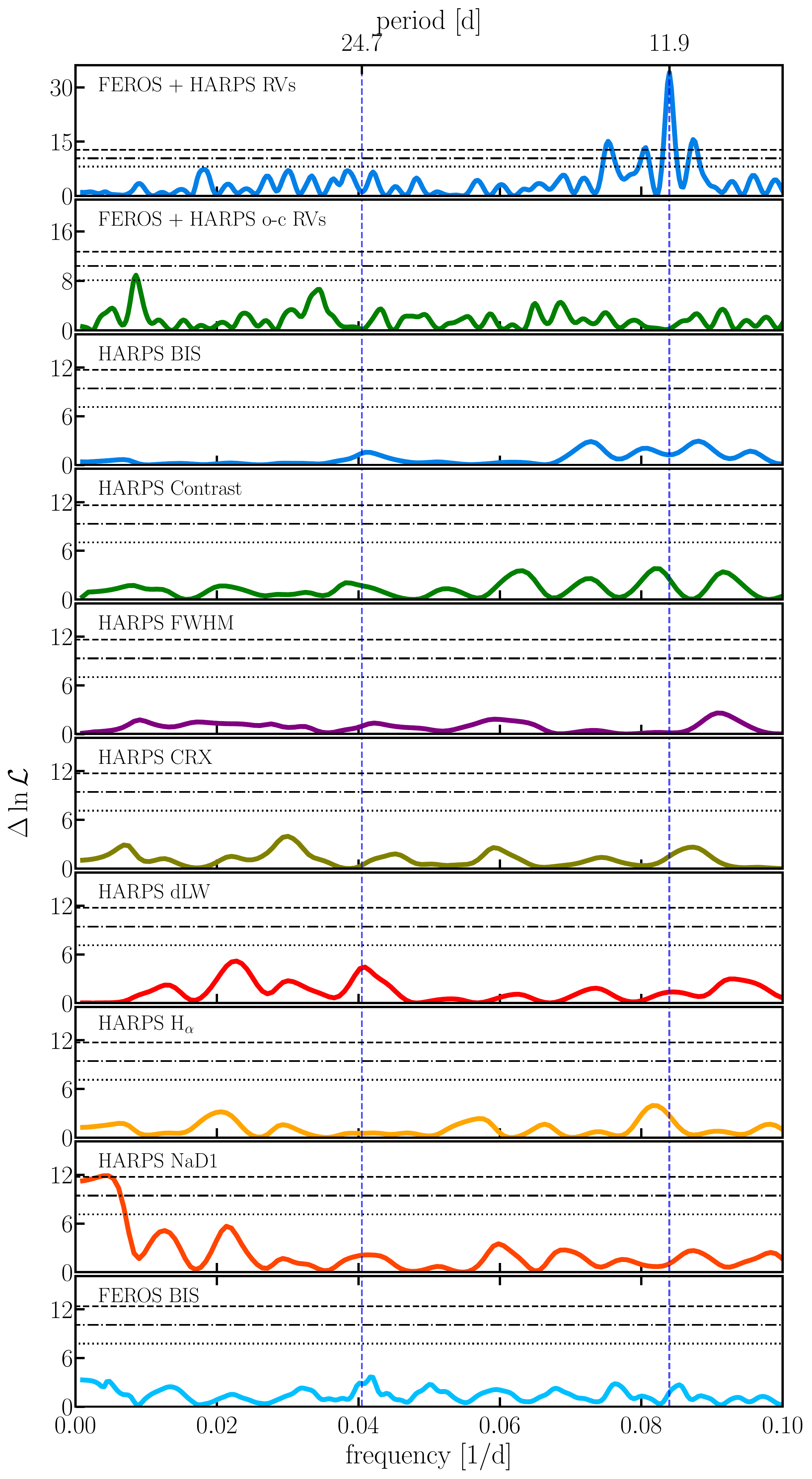} 
 
    \caption{MLP power spectrum for the TOI-2202 data, based on FEROS and HARPS RVs and stellar activity indicators as labeled in the panels. The horizontal lines in the MLP periodograms show the FAP levels 
    of 10\%, 1\%, and 0.1\% in $\Delta\ln\mathcal{L}$.
    Blue vertical lines indicate the orbital period of TOI-2202\,b and c.
    }
    \label{MLP_results} 
\end{figure}

However, a single-planet transit model with a period and phase adopted from the TLS failed to produce a good fit to the available $\tess$ and CHAT data. Performing a transit light curve model optimization by adjusting the orbital
period $P$, 
eccentricity $e$ and 
argument of periastron $\omega$ (or $e \sin\omega$, $e \cos\omega$),
inclination $i$,
time of inferior transit conjunction $t_{0}$,  the semi-major axis relative to the stellar radius $a/R_{\star}$, and planetary radius relative to the stellar radius $R/R_{\star}$, did not help either. 
We found that the light curve transit signals exhibit strong deviations in the expected individual time-of-transits assuming a constant period, suggesting strong transit timing variations (TTVs).
To extract the TTVs, we performed a separate one-planet fit to the $\tess$ and CHAT light curves, 
assuming a circular orbit (i.e. $e \sin\omega$, $e \cos\omega_b$ = 0), but with variable mid-transit times as fitting parameters.
The rest of the transit parameters in this modeling scheme across each individual $\tess$ and CHAT light curve were shared, with exception of the limb-darkening (LD) coefficients of $\tess$ and CHAT, for which we adopted separate quadratic LD models, and nuisance parameters such as the transit light curve relative photometric offset and additional data jitter\footnote{By 'jitter', we mean the unknown variance parameter, which is added in quadrature to the transit photometry and RV error budget. See \citep{Baluev2009}.} To perform an adequate parameter search, we ran a nested sampling scheme with 1000 ``live-points'', focused on the posterior convergence instead of Bayesian evidence \citep[see,][for details]{Speagle2020}.
We adopted the 68.3\,\% confidence levels of the nested sampling posterior probability parameter distribution as $1\sigma$ parameter uncertainties.

The extracted mid-transit time estimates and their precise uncertainties yielded very strong periodic TTVs in the $\tess$ and CHAT data, 
whose amplitude around the mean period was $\sim$ 1.2 hours, covering one full TTV super-period. 
\autoref{table:TTVdata} lists the precisely extracted individual transit times and their errors.
No significant TLS power is detected in the $\tess$ photometry residuals, 
meaning that only one planetary companion of TOI-2202 is detectable on the $\tess$ FFI light curves.
These results suggested the presence of an additional non-transiting 
companion in the TOI-2202 systems that is close and massive enough to perturb the Jovian-like transiting planet TOI-2202 b.

\begin{figure}[tp]
\begin{center}$
\begin{array}{ccc}

\includegraphics[width=9cm]{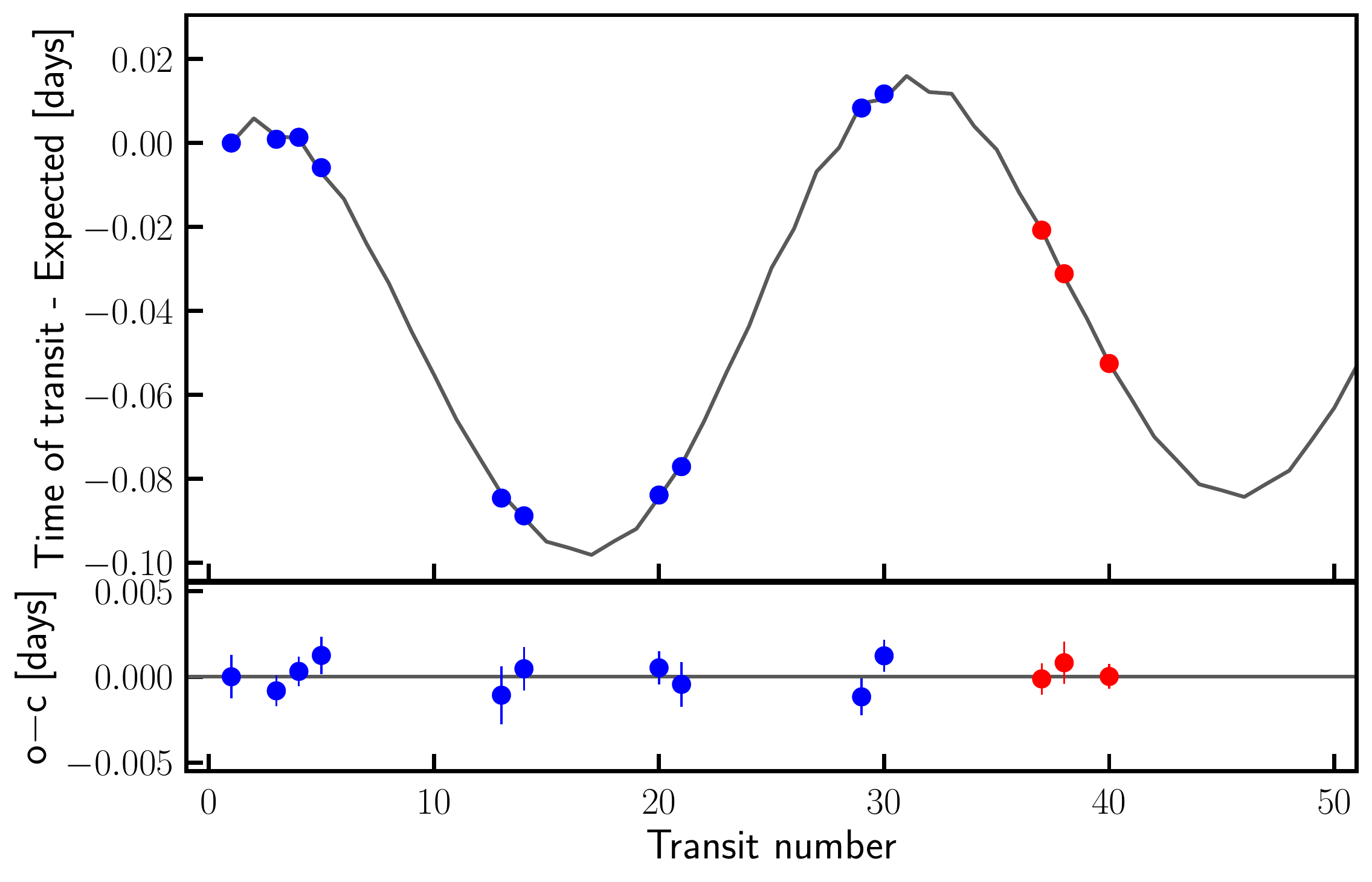} \\
\includegraphics[width=9cm]{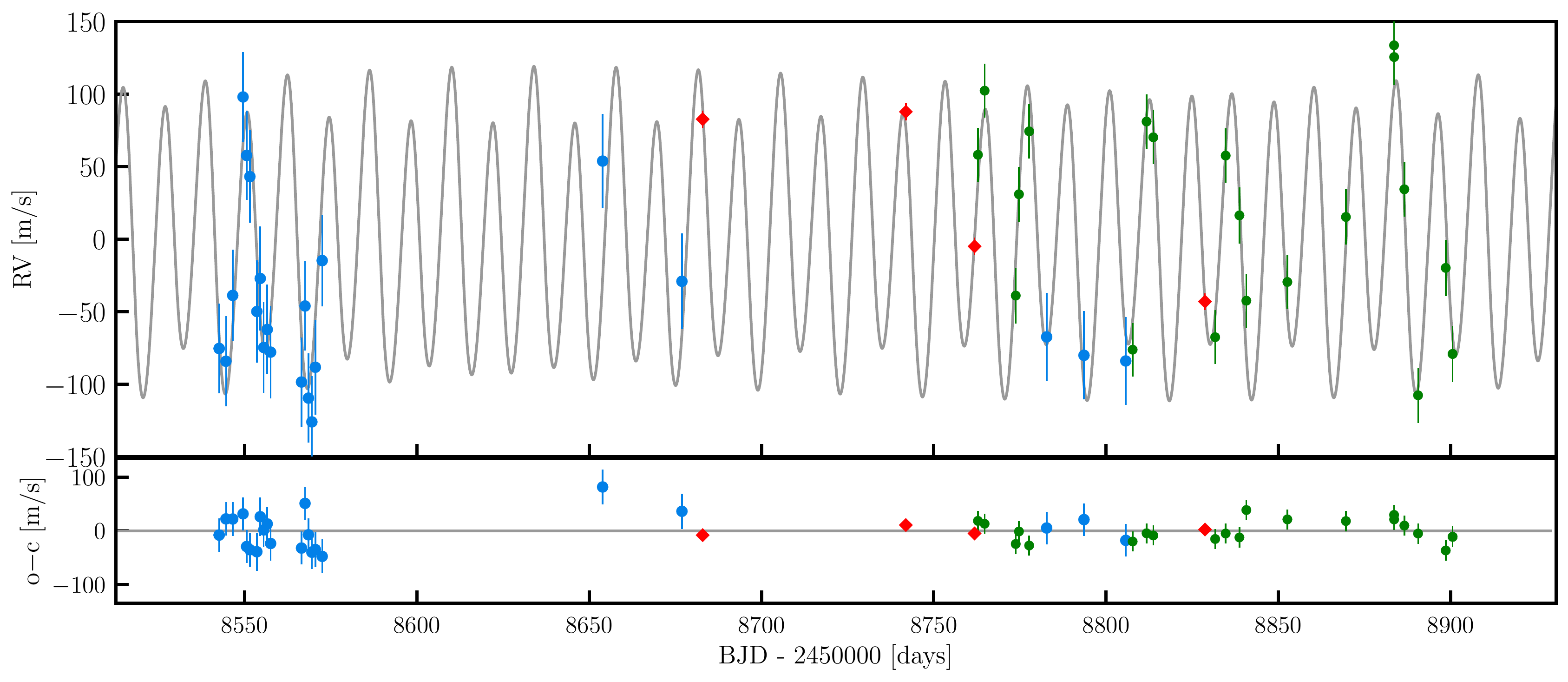}  \\
\includegraphics[width=9cm]{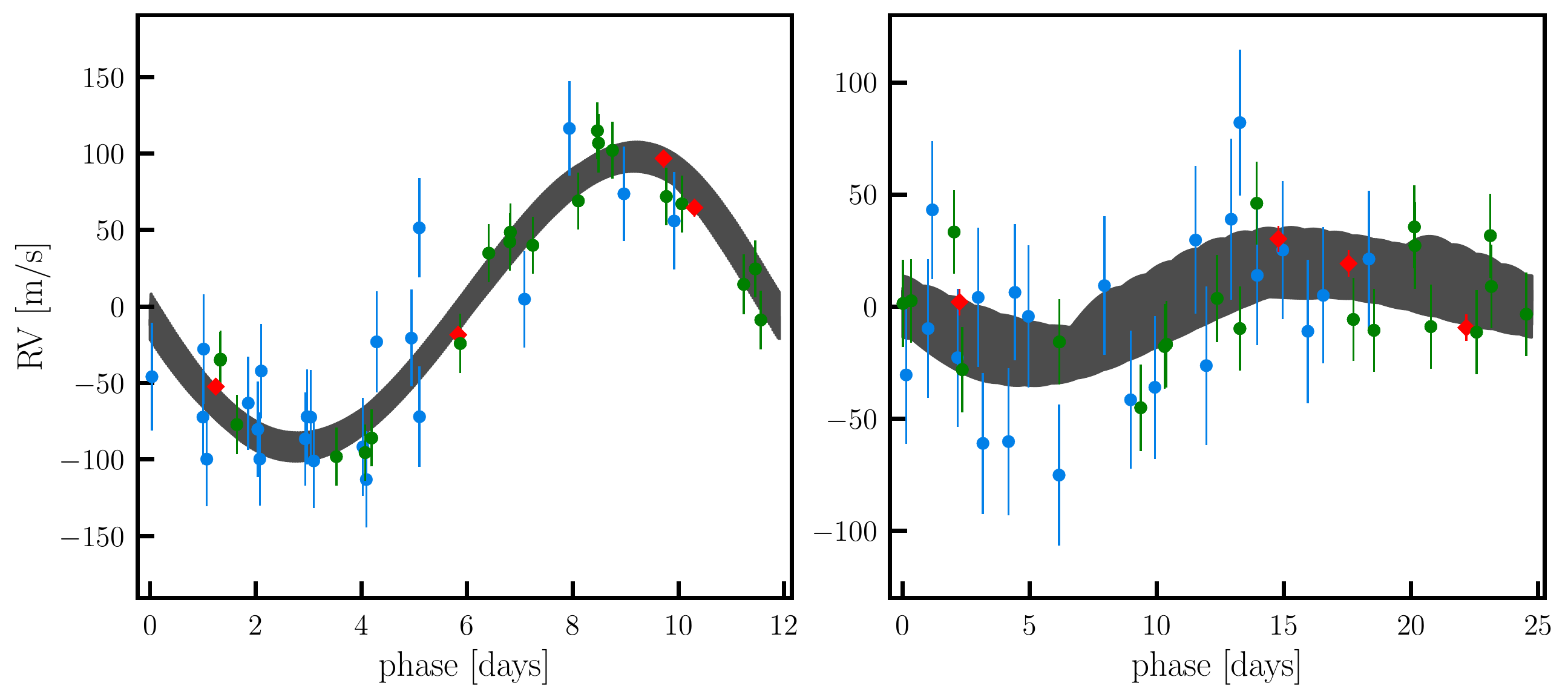}   \\

\end{array} $
\end{center}

\caption{
    TTVs of TOI-2202 ($blue$ -- $\tess$, $red$ --CHAT) modeled with a two-planet dynamical model jointly 
    with the RVs from the FEROS (blue circles), PFS(red diamonds), and HARPS (green circles).
     The top panel shows the TTVs time series and a model consistent with two massive planets with periods
    close to the 2:1 MMR commensurability.  The bottom sub-panel shows the TTVs residuals. 
     The middle panel shows the same model, but for the Doppler data. The bottom sub-panel shows the RVs residuals.
     The bottom left and right panels show a phase-folded representation of the RV data, modeled with 
    the dynamical model (with an osculating period). The data uncertainties include the estimated RV jitter, added in quadrature to the error budget.
}
 
\label{TTV_plot1} 
\end{figure}

\begin{figure}[tp]
\begin{center}$
\begin{array}{ccc}

\includegraphics[width=9cm]{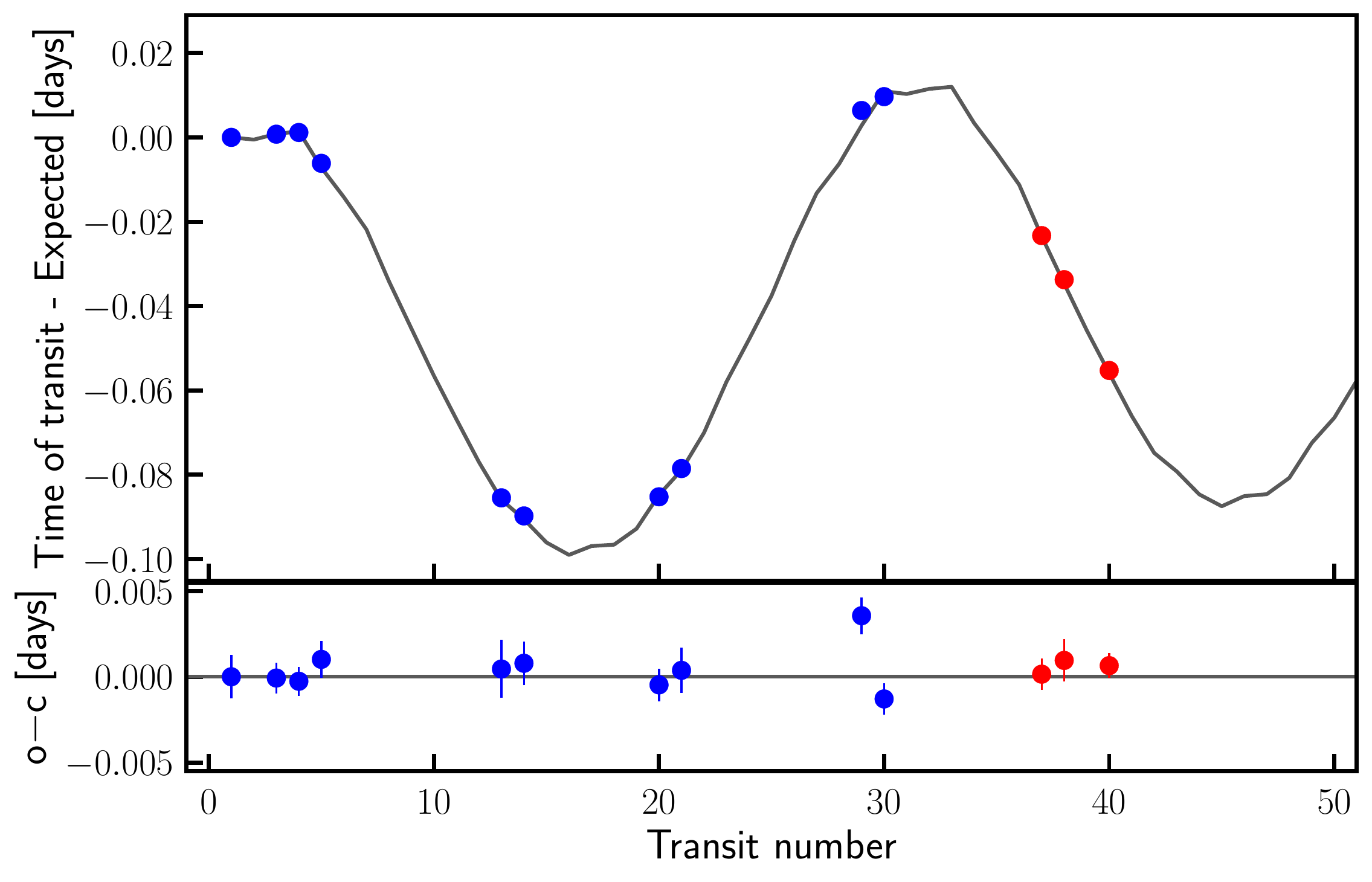}\\
\includegraphics[width=9cm]{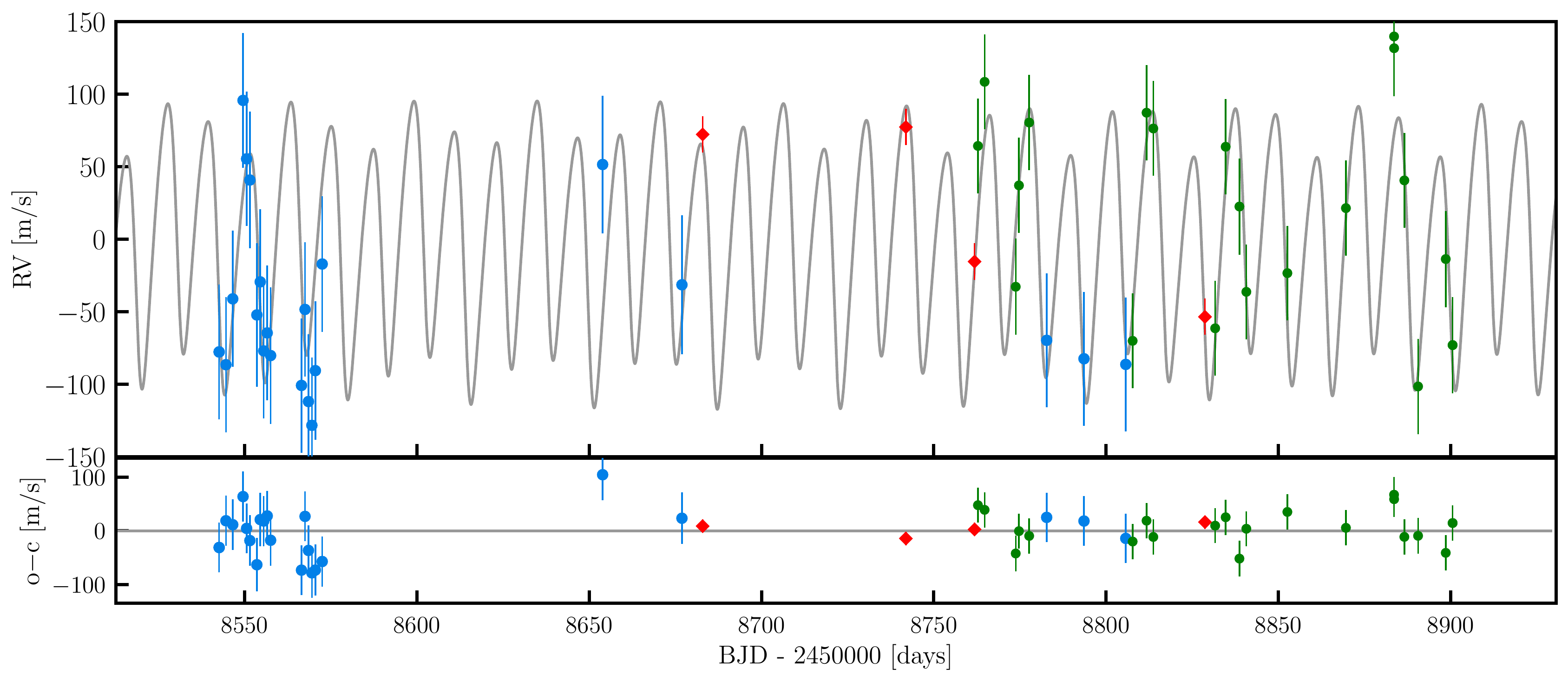} \\
  \includegraphics[width=9cm]{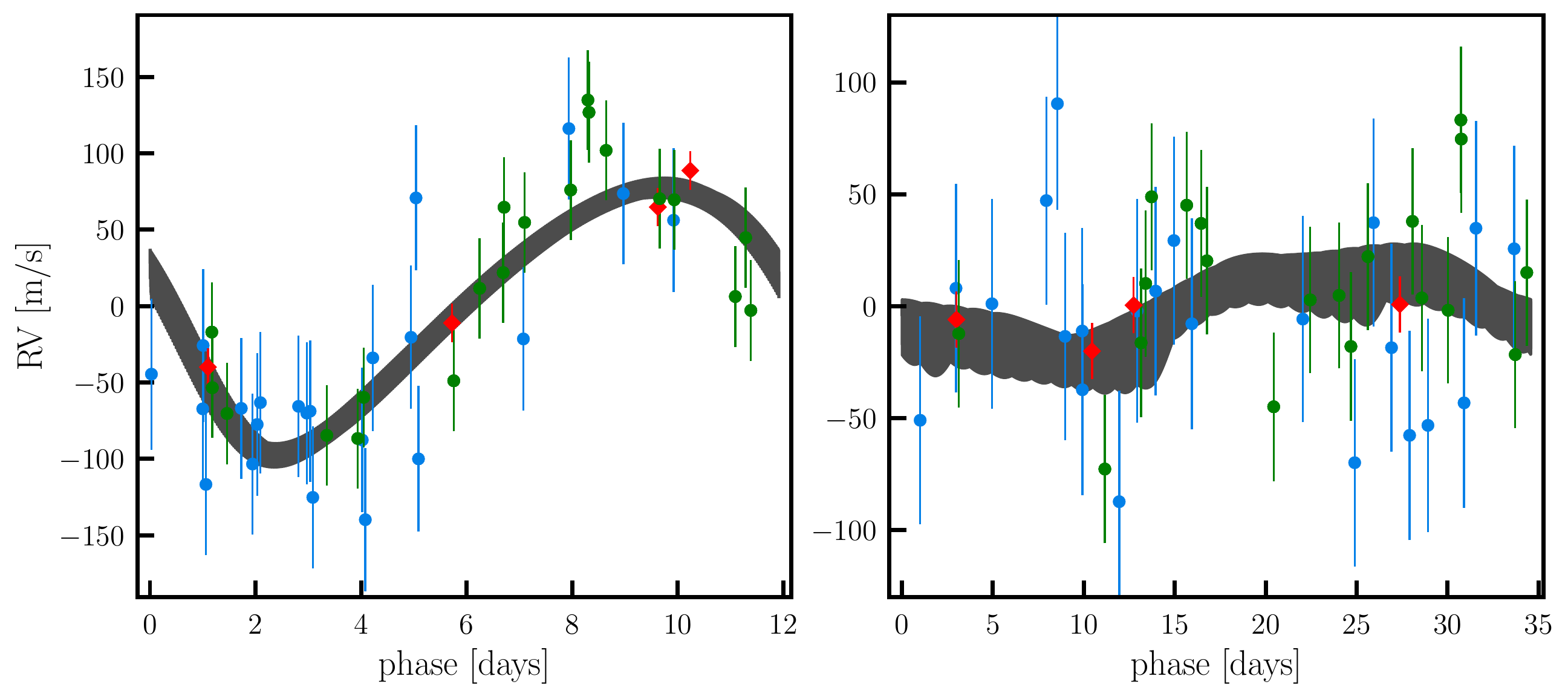}  \\

\end{array} $
\end{center}

\caption{
    Same as Fig.~\ref{TTV_plot1}, but for an alternative dynamical model consistent with two massive planets with periods
    close to the 3:1 MMR commensurability (see text for details).
}
 
\label{TTV_plot2} 
\end{figure}

\begin{table*}[ht]

\centering   
\caption{{\tt Exo-Striker} posteriors probability and maximum $-\ln\mathcal{L}$ orbital parameters estimates of the two-planet system TOI-2202 derived from TTVs extracted from $\tess$ and CHAT light curves and modeled together with precision RV data from FEROS, PFS, and HARPS. Two possible orbital configurations could explain the observed data; a system close to the first order 2:1 MMR, and a system close to the second order 3:1 MMR, with significant statistical preference to the former. } 
\label{TTV_params}

\begin{adjustwidth}{-1.8cm}{}
\resizebox{0.78\textheight}{!}
{\begin{minipage}{1.1\textwidth}

\begin{tabular}{lrrrrrrrrrr}     

\hline\hline  \noalign{\vskip 0.7mm}

\makebox[0.1\textwidth][l]{\hspace{60 mm}$\sim$2:1 MMR fit      \hspace{60 mm} $\sim$3:1 MMR fit  \hspace{1.5 mm} } \\
\makebox[0.1\textwidth][l]{\hspace{40 mm} Median and $1\sigma$  \hspace{15 mm} Max. $-\ln\mathcal{L}$     \hspace{25 mm} Median and $1\sigma$  \hspace{20 mm} Max. $-\ln\mathcal{L}$  \hspace{1.5 mm} } \\
\cline{2-5}\cline{7-10}\noalign{\vskip 0.9mm}

Parameter &\hspace{10.0 mm} Planet b & Planet c & Planet b & Planet c  & & \hspace{10.0 mm}Planet b & Planet c & Planet b & Planet c \\
\cline{1-5}\cline{7-10}\noalign{\vskip 0.7mm}

$K$  [m\,s$^{-1}$]            &  94.2$_{-3.6}^{+3.4}$ & 15.2$_{-2.4}^{+2.6}$ &  95.3 &  19.6 &  
                      &  83.1$_{-7.7}^{+8.0}$ & 22.6$_{-2.7}^{+2.6}$ &  86.9 &  19.5  \\ \noalign{\vskip 0.9mm}

$P$  [day]                    & 11.9108$_{-0.0009}^{+0.0009}$ & 24.7545$_{-0.0078}^{+0.0073}$ & 11.9103 & 24.7557 &  
                      & 11.9145$_{-0.0010}^{+0.0009}$ & 34.6091$_{-0.0125}^{+0.0151}$ & 11.9164 & 34.5803\\ \noalign{\vskip 0.9mm}

$e$                           & 0.0782$_{-0.0097}^{+0.0135}$ & 0.0110$_{-0.0064}^{+0.0094}$ & 0.0672 &  0.0104 &  
                      & 0.1795$_{-0.0172}^{+0.0184}$ & 0.1318$_{-0.0170}^{+0.0210}$ & 0.2069 &  0.1308 \\ \noalign{\vskip 0.9mm}

$\omega$  [deg]               & 26.9$_{-5.6}^{+5.2}$  & 55.7$_{-14.8}^{+16.9}$ & 44.4  & 103.24&   
                      & 109.8$_{-4.5}^{+5.9}$ & 198.9$_{-7.8}^{+7.1}$  & 117.9 & 205.8 \\ \noalign{\vskip 0.9mm}

$M_{\rm 0}$  [deg]            & 71.2$_{-5.5}^{+6.5}$  & 145.8$_{-18.7}^{+13.3}$ & 51.2  & 101.7 &  
                      & 332.3$_{-8.3}^{+6.2}$ &  202.8$_{-4.7}^{+5.6}$  & 319.1 & 200.3 \\ \noalign{\vskip 0.9mm}

$a$  [au]                     &  0.0956$_{-0.0016}^{+0.0015}$  & 0.1558$_{-0.0026}^{+0.0025}$  &      0.0957 & 0.1558 &     
                      &  0.0957$_{-0.0016}^{+0.0015}$  & 0.1948$_{-0.0032}^{+0.0031}$ & 0.0957 & 0.1947 \\ \noalign{\vskip 0.9mm}

$m$  [$M_{\rm jup}$]          & 0.927$_{-0.048}^{+0.047}$  & 0.191$_{-0.030}^{+0.033}$ & 0.939 & 0.246 &     
                      & 0.806$_{-0.081}^{+0.081}$  &           0.316$_{-0.039}^{+0.037}$  &  0.840 & 0.271  \\ \noalign{\vskip 0.9mm}
 
\cline{1-5}\cline{7-10}\noalign{\vskip 0.7mm}   

RV$_{\rm off}$ FEROS  [m\,s$^{-1}$]               &        -126.5$_{-7.1}^{+7.4}$  & & -125.9 & &   &      -128.4$_{-7.3}^{+8.9}$  & & -123.5 \\ \noalign{\vskip 0.9mm}
RV$_{\rm off}$ PFS  [m\,s$^{-1}$]                &        -76.3$_{-3.7}^{+3.6}$   & & -79.5 &  &   &      -65.5$_{-11.7}^{+9.1}$  & & -67.0  \\ \noalign{\vskip 0.9mm}
RV$_{\rm off}$ HARPS  [m\,s$^{-1}$]               &        -109.7$_{-5.1}^{+4.7}$  & & -110.5 & &   &      -111.6$_{-7.2}^{+6.2}$  & & -116.6  \\ \noalign{\vskip 0.9mm}
RV$_{\rm jit}$ FEROS  [m\,s$^{-1}$]               &          30.7$_{-5.5}^{+7.3}$  & &  29.0 & &    &      44.3$_{-6.5}^{+8.4}$    & & 45.2  \\ \noalign{\vskip 0.9mm}
RV$_{\rm jit}$ PFS  [m\,s$^{-1}$]                &           5.9$_{-5.7}^{+7.0}$  & & 5.4  & &     &      20.6$_{-6.7}^{+12.3}$   & &  12.3 \\ \noalign{\vskip 0.9mm}
RV$_{\rm jit}$ HARPS  [m\,s$^{-1}$]               &           19.4$_{-3.8}^{+4.6}$ & & 17.4 & &     &      31.1$_{-5.1}^{+7.1}$    & & 32.0 \\ \noalign{\vskip 0.9mm}

$-\ln\mathcal{L}$             &       & &  -136.448  &&   &  & &        -159.336 \\
\\
\hline \noalign{\vskip 0.7mm}

\end{tabular}

\end{minipage}}
\end{adjustwidth}
\tablecomments{The orbital elements are in the Jacobi frame and are valid for epoch BJD = 2458327.103. Only co-planar and edge-on systems ($i_b$,$i_c$~=~90$^{\circ}$ and $\Delta i$ = 0$^{\circ}$) are assumed. The joint TTV+RV dynamical model accepts a fixed value of the stellar mass (0.823 $M_{\odot}$); however, the derived planetary posterior parameters of $a$, and $m$ are calculated taking into account the stellar mass uncertainty according to the floor uncertainties predicted by \citet{Tayar2020}, listed in \autoref{table:phys_param}.}

\end{table*}


\subsection{Spectral RV and activity indices analysis}
\label{Sec4.3}

For period search in the precise RVs and activity indices data of TOI-2202, we computed 
maximum likelihood periodograms, which calculate the log-likelihood ($\ln\mathcal{L}$) power by optimizing for each test frequency. 
The MLP algorithm allows for multiple data sets, each with an additive offset and jitter parameters \citep{Baluev2009, Zechmeister2019}, which makes it more suitable for period analysis of multi-instrument data.
We adopted the significance thresholds of the $\Delta\ln\mathcal{L}$ improvements, 
which correspond to false-alarm probabilities (FAPs) of 10\,\%, 1\,\%, and 0.1\,\%.

\autoref{MLP_results} shows the MLP periodograms of the combined FEROS and HARPS RVs and activity time series, separately. The PFS data consist of only four RVs and these were not included in the MLP analysis.
The MLP periodogram of the combined FEROS and HARPS data shows a strong power at 11.91\,d, which is the period of the known transiting planet candidate. 
The RV residuals, after removing the 11.9\,d signal, however, do not indicate the presence of additional significant periodicity in the RV data.
No significant activity periodicity is evident in the FEROS and HARPS activity indices. The only exception is 
the HARPS Na D1 activity index data, which show marginally significant $\Delta\ln\mathcal{L}$  at lower frequencies, which, however, do not have 
a counterpart in the HARPS RVs. 
Thus, we concluded that the MLP analysis of the activity data alone does not suggest that TOI-2202 is an active star.

Further RV analysis in this work was performed jointly with just the available TTVs (\autoref{Sec4.4}) and transit photometry data (\autoref{Sec4.5}) 
of TOI-2202, using dynamical modeling.

\begin{figure*}[tp]
    \centering
    \includegraphics[width=8.9cm]{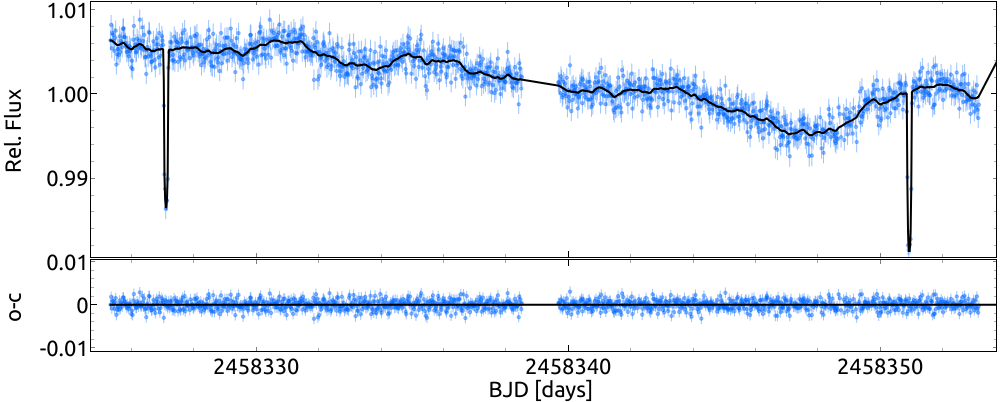} \put(-20,90){a)}
    \includegraphics[width=8.9cm]{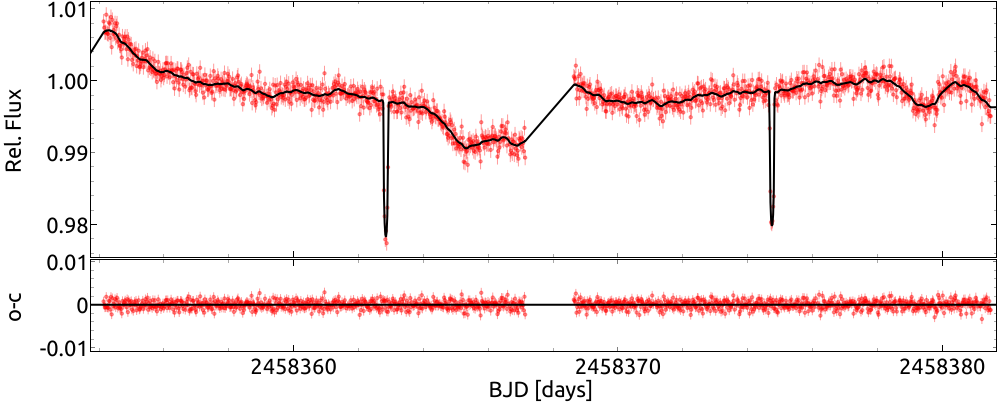} \put(-20,90){b)} \\ 
    \includegraphics[width=8.9cm]{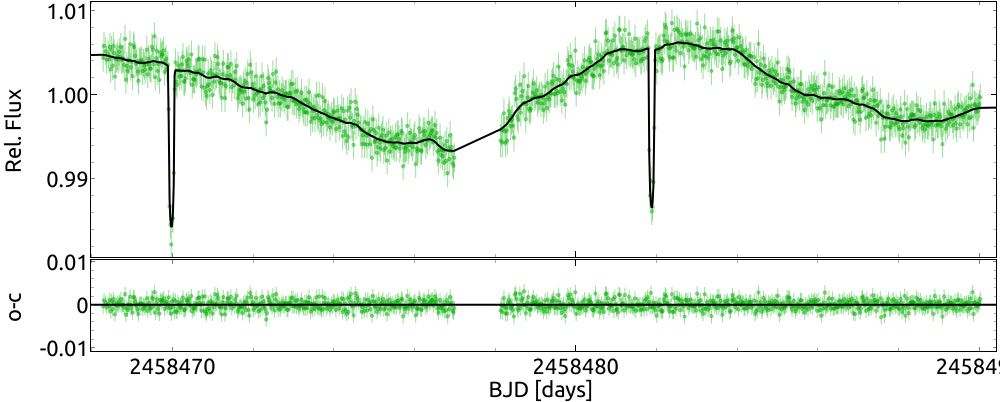} \put(-20,90){c)}
    \includegraphics[width=8.9cm]{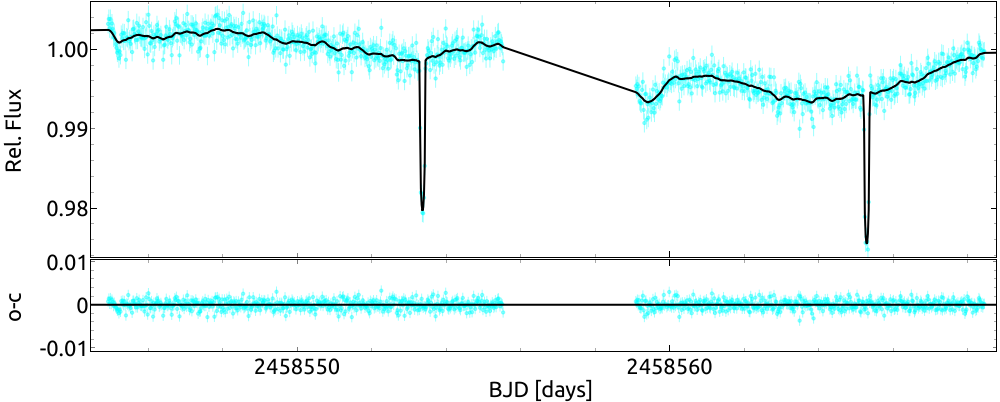} \put(-20,90){d)} \\  
    \includegraphics[width=8.9cm]{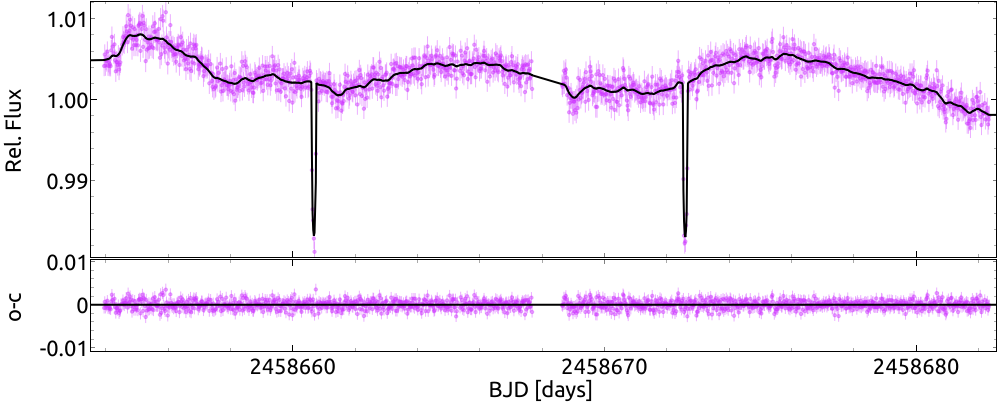}  \put(-20,90){e)}  
    \includegraphics[width=8.9cm]{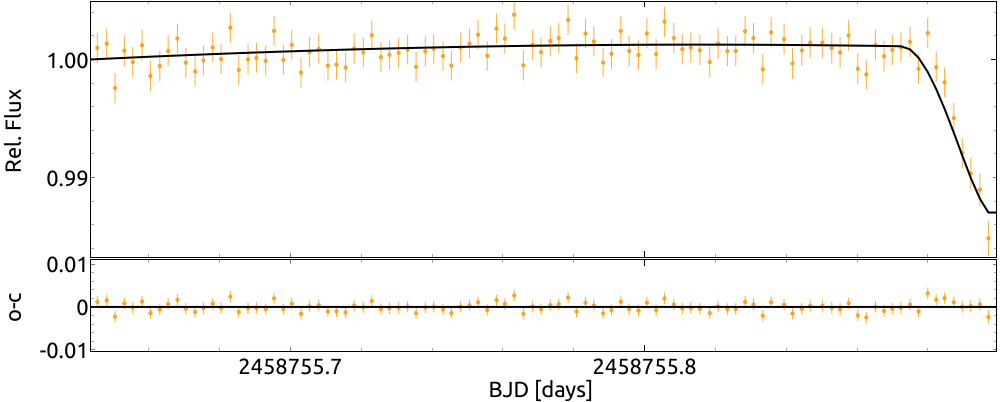} \put(-20,90){f)} \\    
    \includegraphics[width=8.9cm]{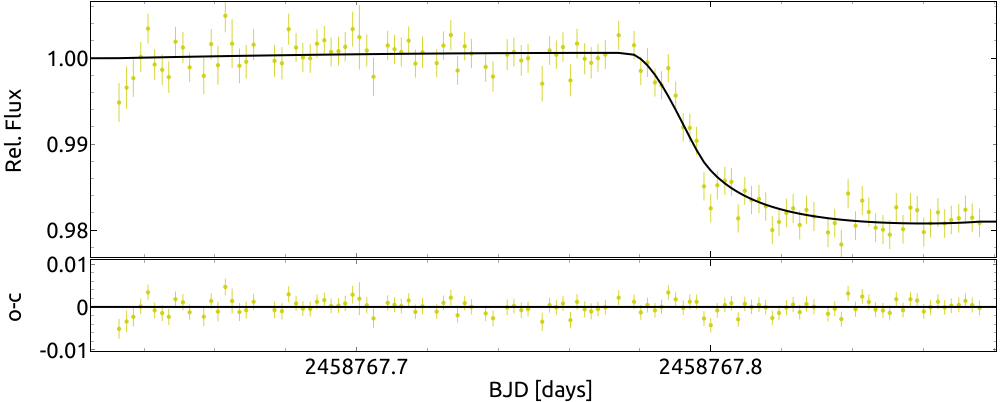} \put(-20,90){g)}  
    \includegraphics[width=8.9cm]{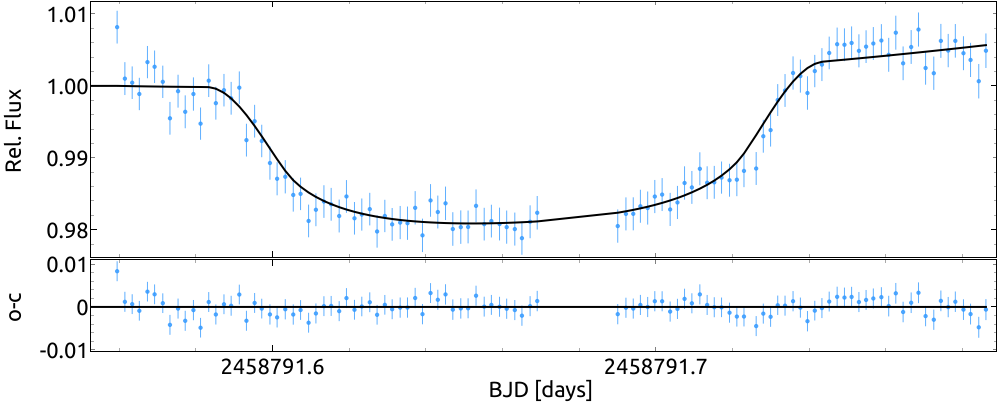} \put(-20,73){h)} \\

\caption{Panels a)-e) show TOI-2202's $\tess$ FFI raw photometry data reduced with {\tt tesseract} from Sectors: 1 (blue), 2 (red), 6 (green), 9 (orange) and 13 (magenta). Panels f)-h) show the three transit events recorded with CHAT. The black curve on panels shows the global photo-dynamical model constructed together with the RV model of FEROS, HARPS, and PFS, including a common transit light curve and RV Gaussian processes regression model that serves as a proxy of the stellar activity (see \autoref{RV_data_GP}). 
The CHAT light curves are de-trended against airmass using linear models simultaneously fitted with the rest of the orbital and nuisance data parameters.  The subplots of panels a)-e) show the residuals between the model (black line) and the respective light curve data.
}
 
\label{tra_data} 
\end{figure*}

\begin{figure*}[tp]
    \centering
    \includegraphics[width=17.5cm]{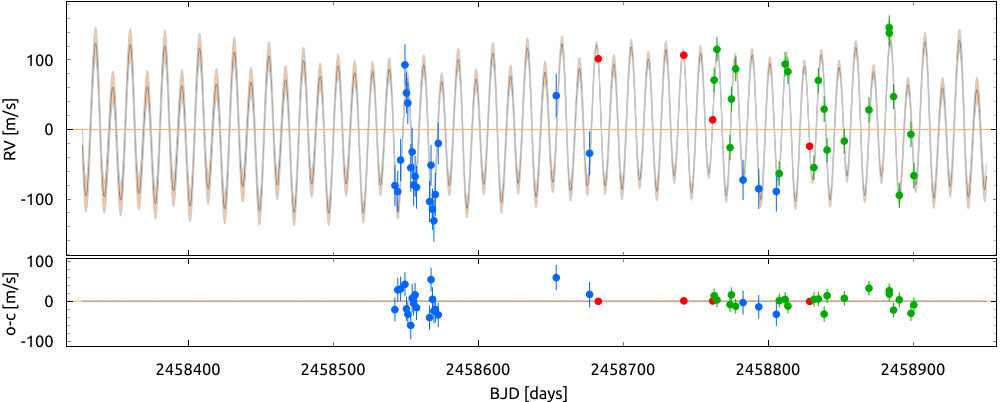} \\
    \includegraphics[width=17.5cm]{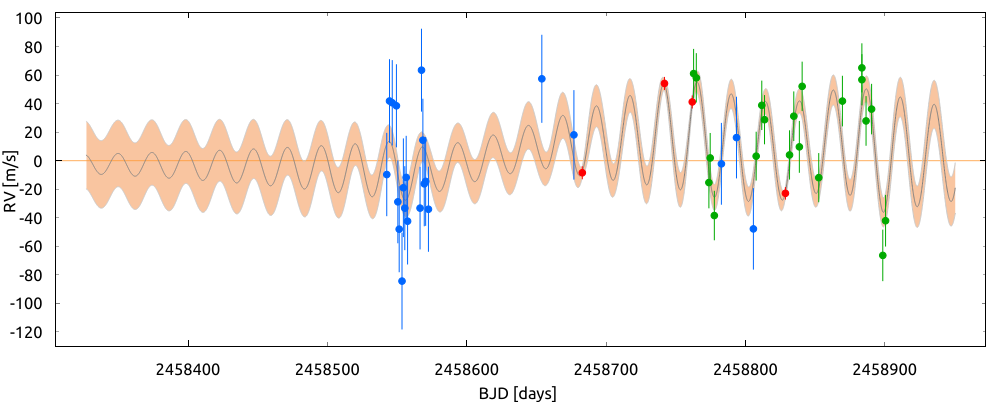}  \\ 

\caption{The top panel shows the TOI-2202 RV data from FEROS (blue), HARPS (green), and PFS (red) modeled with a global dynamical model together with the transit light curves of $\tess$ and CHAT including a common RV and transit GP regression model. The subplot to the top panel shows the residuals between the model (black line) and the RV data. The bottom panel shows the GP model component after subtracting the RV counterpart of the dynamical model. 
}
 
\label{RV_data_GP} 
\end{figure*}

\subsection{Joint TTVs and RV analysis}
\label{Sec4.4}

Logically, our next step was to study the TTVs obtained from the $\tess$ and CHAT light curve transit events.
TTVs contain essential information of the system's dynamics  \citep{Agol2005}, and adequate modeling could reveal the most likely orbital parameters and planetary masses.
This is under the condition that at 
least one complete TTV super-period (i.e., one cycle of the TTV signal) is covered  \citep{Lithwick2012}. We detected over one full TTV super-period 
of the transiting TOI-2202 b, with an estimated period of approximately 357 days.
Although this is a promising lead, following \citet{Lithwick2012}, it is evident 
that the observed super-period and the TTVs amplitude can 
be explained with almost all possible period ratios in the low-order first or second order commensurability, in which TOI-2202 c could be inner or outer planet. Therefore, we examined wide ranges of the period, mass and eccentricity of the non-transiting TOI-2202 c.

First, we performed TTV fitting with the {\tt Exo-Striker} by adopting a self-consistent dynamical model on the extracted TTVs. 
The fitted parameters for each planet were the dynamical planetary mass $m$, orbital period~$P$, eccentricity $e$, argument of periastron $\omega$, and mean-anomaly $M_0$, which in this work are always valid for the epoch of the first transit event $t_0$ = 2458327.103 [BJD]. For this test we assumed coplanar, edge-on and prograde 
two-planet system (i.e., $i_b$,$i_c$ = 90$^\circ$ and $\Delta i$ = 0$^\circ$),
 whereas for the dynamical mass of TOI-2202, we adopt our best estimate of 0.823 $M_\odot$.
The time step in the dynamical model was set to 0.1 days to assure
a precise orbital resolution of the inner transiting planet TOI-2202 b.

We ran a nested sampling scheme, which allowed us to efficiently explore the complex parameter space of osculating orbital parameters and study the parameter posteriors. In our nested sampling scheme with {\tt dynesty}, we ran  100 ``live-points'' per fitted parameter, focused on the Bayesian log-evidence convergence, using a ``static" nested sampler \citep[see,][for details]{Speagle2020}.  
For all parameters, we adopt uniform priors, which define an equal probability of occurrence within the experimentally chosen parameter ranges. For TOI-2202 b, we select prior parameter range estimates
 taken from our single-planet parameter analysis from \autoref{Sec4.2}, which assure transit occurrence near $t_0$ = 2458327.103, whereas for the perturber in the system (to become TOI-2202 c), we explore a wide parameter space of eccentricities, masses and periods. For instance, 
we explore $P_c \in \mathcal{U}$(20.0,40.0)\,d, $e_c \in \mathcal{U}$(0.0,0.2), and $m_c \in \mathcal{U}$(0.01,0.6) $M_{\rm jup.}$; the rest of the priors shall not be discussed here for the sake of brevity. However, the prior ranges can be visually assessed in \autoref{Nest_samp_ttv}, which shows the resulted posterior probability distribution from the TTV analysis. \autoref{Nest_samp_ttv} indicates that the posteriors are multi-modal, suggesting that more than one orbital solutions could explain the data.

We find that many pairs of planetary masses and eccentricities, and orbital periods of TOI-2202 c provide a plausible explanation of the extracted TTVs of TOI-2202 b.
The ambiguity in eccentricity versus dynamical planetary mass 
was already observed in another K-dwarf TESS system consistent with two warm Jovian-mass planets, TOI-216 \citep{Dawson2019}. Based only on the TTVs for this system, \citet{Dawson2019} were unable to firmly decide whether the system is an eccentric pair of a Saturn-mass planet accompanied by a Neptune-mass planet or is composed of a Jovian-mass planet in a 2:1 MMR with a sub-Saturn-mass planet, where both planets are consistent with more circular orbits. 
Only after securing a large number of RV measurements and expanding the TTVs baseline of TOI-216 \citep{Dawson2021} confirmed the latter configuration.
From the posterior probability distribution, we find that TTVs of TOI-2202\,b are most likely induced by an exterior sub-Saturn, whose orbital period is close to the first-order eccentricity-type 2:1 MMR with the transiting planet. Such a planetary configuration resembles the solution of TOI-216. Nevertheless, as can be seen from \autoref{Nest_samp_ttv}, at this stage we cannot completely rule out that the TOI-2202 system is more eccentric and resides in the second-order MMR in the 3:1 period ratio commensurability.

The precise RV measurements that we obtained for TOI-2202 could further constrain the orbital eccentricity and planetary masses, and break the ambiguity. Therefore, as a next step, we included the RV data in the analysis by performing a joint TTV+RV  nested sampling analysis, repeating the steps and prior ranges listed above. The RV inclusion in the modeling leads to a more complex model, which now fits the RV semi-amplitude $K$ parameter for each planet, constraining the planetary masses, and the RV data offsets and RV jitter parameters for HARPS, PFS and FEROS, adding six more free parameters. For these, we defined experimentally defined uniform priors of RV off. $ \in \mathcal{U}$(--140.0,--70.0) m\,s$^{-1}$, and log-uniform (Jeffreys) priors of RV jitter $\in \mathcal{J}$(0.0,50.0) m\,s$^{-1}$.

\begin{table*}[ht]

\centering   
\caption{{Nested sampling priors, posteriors, and maximum $-\ln\mathcal{L}$ orbital parameters of the two-planet system TOI-2202 derived by 
joint dynamical modeling of photometry ($\tess$, CHAT) and radial velocities (FEROS, PFS, HARPS). Additionally, these parameters were estimated including a Gaussian processes regression kernel, which was used to model the stellar rotation effects and was common to the $\tess$ light curves and RV data. CHAT light curves were simultaneously modeled with linear regression models to account for air mass optical effects. These, and other nuisance parameter estimates are listed in \autoref{NS_params2}. }}
\label{NS_params}

 \begin{adjustwidth}{-3cm}{}
 \resizebox{0.92\textheight}{!}
 {\begin{minipage}{1.1\textwidth}

\begin{tabular}{lrrrrrrrrrrrr}     

\hline\hline  \noalign{\vskip 0.7mm}

\makebox[0.1\textwidth][l]{\hspace{40 mm} Median and $1\sigma$  \hspace{20 mm} Max. $-\ln\mathcal{L}$     \hspace{20 mm} Adopted priors  \hspace{10 mm} \hspace{1.5 mm} } \\
\cline{1-9}\noalign{\vskip 0.7mm}

Parameter &\hspace{10.0 mm} Planet b & Planet c &  & Planet b & Planet c  & & \hspace{10.0 mm}Planet b & Planet c  \\
\cline{1-9}\noalign{\vskip 0.7mm}





$K$  [m\,s$^{-1}$]            &  99.2$_{-5.2}^{+5.6}$ & 29.3$_{-6.6}^{+8.3}$ &  
                              &  92.9 & 38.3 &
                              &  $\mathcal{U}$(90.0,110.00) & $\mathcal{U}$(20.0,40.0)  &  \\ \noalign{\vskip 0.9mm}

$P$  [day]                    & 11.9101$_{-0.0036}^{+0.0022}$ & 24.6744$_{-0.0339}^{+0.0258}$ & 
                              & 11.9123 & 24.6797 &  
                              & $\mathcal{U}$(11.90,11.92) &  $\mathcal{U}$(24.60,24.80)  &  \\ \noalign{\vskip 0.9mm}

$e$                           & 0.0420$_{-0.0075}^{+0.0255}$ & 0.0622$_{-0.0211}^{+0.0452}$ &
                              & 0.0672 &  0.0104 &  
                              & $\mathcal{U}$(0.00,0.2) &  $\mathcal{U}$(0.00,0.2)  &  \\ \noalign{\vskip 0.9mm}

$\omega$  [deg]               & 84.1$_{-16.4}^{+9.8}$  & 320.8$_{-12.5}^{+129.7}$ &
                              & 86.0  &  322.6 &   
                              & $\mathcal{U}$(0.0,360.00) &  $\mathcal{U}$(0.0,360.00) &  \\ \noalign{\vskip 0.9mm}

$M_{\rm 0}$  [deg]            & 7.6$_{-8.5}^{+26.4}$  & 267.2$_{-269.9}^{+9.6}$ &
                              & 4.3  & 265.8 &  
                              & $\mathcal{U}$(0.0,360.00) &  $\mathcal{U}$(0.0,360.00)  &  \\ \noalign{\vskip 0.9mm}

$\lambda$  [deg]          &  90.4$_{-0.8}^{+14.6}$  & 226.4$_{-15.5}^{+7.1}$ & 
                              & 90.3 & 228.4 &     
                              &  (derived) &   (derived)  &  \\ \noalign{\vskip 0.9mm}



$i$          [deg]            & 88.4$_{-3.3}^{+0.6}$  & 84.7$_{-2.9}^{+2.4}$ &
                              & 88.9  & 87.2 &  
                              & $\mathcal{U}$(80.0,90.0) &  $\mathcal{U}$(80.0,90.0)  &  \\ \noalign{\vskip 0.9mm}  
                              
$\Omega$     [deg]            & 0.0   & 5.0$_{-2.8}^{+ 2.3}$ &
                              & 0.0  & 5.9 &  
                              & (fixed) &  $\mathcal{U}$(0.0,10.0)  &  \\ \noalign{\vskip 0.9mm}                               

a/$R_\star$                   & 23.08$_{-2.18}^{+2.90}$  & $\dots$ &
                              & 23.45 & $\dots$ &  
                              &  $\mathcal{U}$(20.0,35.00) & $\dots$   &  \\ \noalign{\vskip 0.9mm}
                              
R/$R_\star$                   & 0.1261$_{-0.0065}^{+0.068}$  & $\dots$ &
                              & 0.1255 & $\dots$ &  
                              &  $\mathcal{U}$(0.01,0.25) & $\dots$   &  \\ \noalign{\vskip 0.9mm}
                              
$\Delta i$  [deg]             &  6.56$_{-2.10}^{+1.92}$  & $\dots$ &  
                              &  6.14 & $\dots$ &     
                              &  (derived) &   $\dots$  &  \\ \noalign{\vskip 0.9mm}


$a$  [au]                     &  0.09564$_{-0.00161}^{+0.00156}$  & 0.15544$_{-0.00263}^{+0.00255}$  &  
                              &  0.09569 & 0.15554&     
                              &  (derived) &   (derived)  &  \\ \noalign{\vskip 0.9mm}

$m$  [$M_{\rm jup}$]          &  0.978$_{-0.0588}^{+0.0630}$  & 0.369$_{-0.0836}^{+0.103}$ & 
                              & 0.917 & 0.482 &     
                              &  (derived) &   (derived)  &  \\ \noalign{\vskip 0.9mm}

$R$  [$R_{\rm jup}$]          & 1.01$_{-0.080}^{+0.522}$  & $\dots$ & 
                              & 0.992 & $\dots$ &     
                              &  (derived) &   $\dots$  &  \\ \noalign{\vskip 0.9mm}

\\
\hline \noalign{\vskip 0.7mm}

\end{tabular}

\end{minipage}}
\end{adjustwidth}
\tablecomments{The orbital elements are in the Jacobi frame and are valid for epoch BJD = 2458327.103.  The joint dynamical model accepts a fixed value of the stellar mass (0.823 $M_{\odot}$), however, the derived planetary posterior parameters of $a$, $m$, and $R$ are calculated taking into account the stellar parameter uncertainties (see Note in \autoref{table:TTVdata}. The median value of $m_c$ comes from bimodal distribution (see \autoref{dyn_samp}).  
}
\end{table*}

\autoref{Nest_samp_ttv2} shows the results from this analysis. The posterior probability distribution of the parameters from this test is definitive, suggesting a Jovian-Saturn pair close to the 2:1 MMR. Yet,  some fraction of the samples are samples are consistent with configurations at the 3:1 period ratio commensurability. 
\autoref{TTV_plot1} shows the TTV+RV dynamical fit with maximum $-\ln\mathcal{L}$ from the nested sampling posteriors, which is consistent with a pair of massive planets close to the 2:1 MMR.  For completeness, we examined the alternative 
best fit model near the 3:1 commensurability.
\autoref{TTV_plot2} shows the competing $\sim$3:1 period ratio best-fit solution.  The quality of the these fits and 
overall posterior probability median values and $1\sigma$ uncertainties are listed in \autoref{TTV_params}.
From \autoref{TTV_plot1}  and \autoref{TTV_plot2} it is very clear that a pair close to the 2:1 MMR, or the 3:1 MMR can explain the TTV signal. However, the only reasonable solution to the RV data is shown in \autoref{TTV_plot1}, i.e., the system in the 2:1 commensurability. 
We note that in both cases shown in \autoref{TTV_plot1} and \autoref{TTV_plot2} the RV scatter is large, which we attribute to the relatively low S/N of the spectra that we achieved for this rather faint star, but the RV jitter parameter estimates of the HARPS, FEROS, and PFS data are far more reasonable in the 2:1 period ratio case. A visual inspection of \autoref{TTV_plot1} and \autoref{TTV_plot2} shows that the RV data follows the two-planet dynamical model adequately.  
The orbital solution close to the 2:1 MMR has $-\ln\mathcal{L}$ = -136.45, which is statistically more significant than the one close to the 3:1 MMR, which has $-\ln\mathcal{L}$ = -159.34
(i.e., $\Delta\ln\mathcal{L}$= 22.9).  
Therefore, we conclude that the combination of TTVs from TESS and CHAT, and the RV data from FEROS, HARPS and PFS firmly point to a massive pair of Jovian planets with periods of $P_b$ = 11.9101$_{-0.0009}^{+0.0009}$ days, and P$_c$ = 24.754$_{-0.008}^{+0.007}$ days, 
eccentricities of  $e_b$ = 0.078$_{-0.010}^{+0.014}$ and $e_c$ = 0.011$_{-0.006}^{+0.009}$ and dynamical masses of 
$m_b$ = 0.927$_{-0.036}^{+0.035}$ $M_{\rm jup}$   and $m_c$ = 0.191$_{-0.030}^{+0.032}$ $M_{\rm jup}$.

\subsection{Joint photo-dynamical analysis of the transit light curves and RV data}
\label{Sec4.5}

The transit light curves and the RV data of TOI-2202 contain the dynamical signature of the gravitationally interacting planets in the system. 
Therefore, we performed an alternative, more complex, orbital fitting with respect to the analysis presented in \autoref{Sec4.4}. We adopted a self- consistent photo-dynamical model that fits directly the transit light curves and RVs in an attempt to extract more accurate estimates of the planetary orbital and physical parameters. This comes at the cost of significantly more CPU-time.

We chose to apply the photo-dynamical model on the raw photometric light curves of TOI-2202, which we simultaneously de-trended during the orbital fitting. We included linear models fit to the CHAT data, which simultaneously de-trend the light curves against airmass at the time of observation, and a GP regression model fit to the $\tess$ transit light curves, which aims at capturing the evident stellar activity signals seen in the $\tess$ data (see \autoref{Sec3.1} and \autoref{fig2}). The transit GP model parameters we share with a GP model applied to the RV time series.
The inclusion of a complex transit+RV GP model component to the already complex photo-dynamical model is well justified, since the $\tess$ light curves exhibit periodicity near the estimated orbital period of TOI-2202 c, which could possibly affect its the RV signature, and thereafter, our mass and eccentricity estimates of the planets. 
Similar is the case of the GJ\,143 system \citep[TOI-186,][]{Trifonov2019a,Dragomir2019}, for which the $\tess$ light curves and spectroscopy data are consistent with stellar activity periodicity that is very close to the orbital period of the transiting Neptune-mass planet GJ\,143 b \citep{Gan2020}. 
Although we did not detect significant periodic signals in the spectroscopic activity indices and RVs near the orbital frequency TOI-2202 c, the large RV jitter observed in all three RV data sets motivated us to adopt a ``common" GP regression model. For the purpose, we adopted the rotational GP regression kernel as formulated by \citet{celerite}.

\begin{equation}
 k(\tau) = \frac{B}{2+C}e^{-\tau/L}  \left[ \cos\bigg( \frac{2\pi\tau}{P_{rot}} \bigg) + (1+C) \right] ,
\label{eq:hill1}
\end{equation}

\noindent
where P$_{rot}$ is a proxy for the rotation period of the star, $L$ is the coherence timescale (e.g. life-time of stellar spots), $\tau$ is the time-lag between two consecutive data points, and $C$ is a balance parameter for the periodic and the non-periodic parts of the GP kernel. These parameters were set common to the RV and transit lightcurve parts of the model. The parameter $B$ defines the GP co-variance amplitude, thus naturally, the RV and the transit models have separate amplitude parameters.

\begin{figure*}[tp]
\begin{center}$
\begin{array}{ccc}

\includegraphics[width=15.8cm]{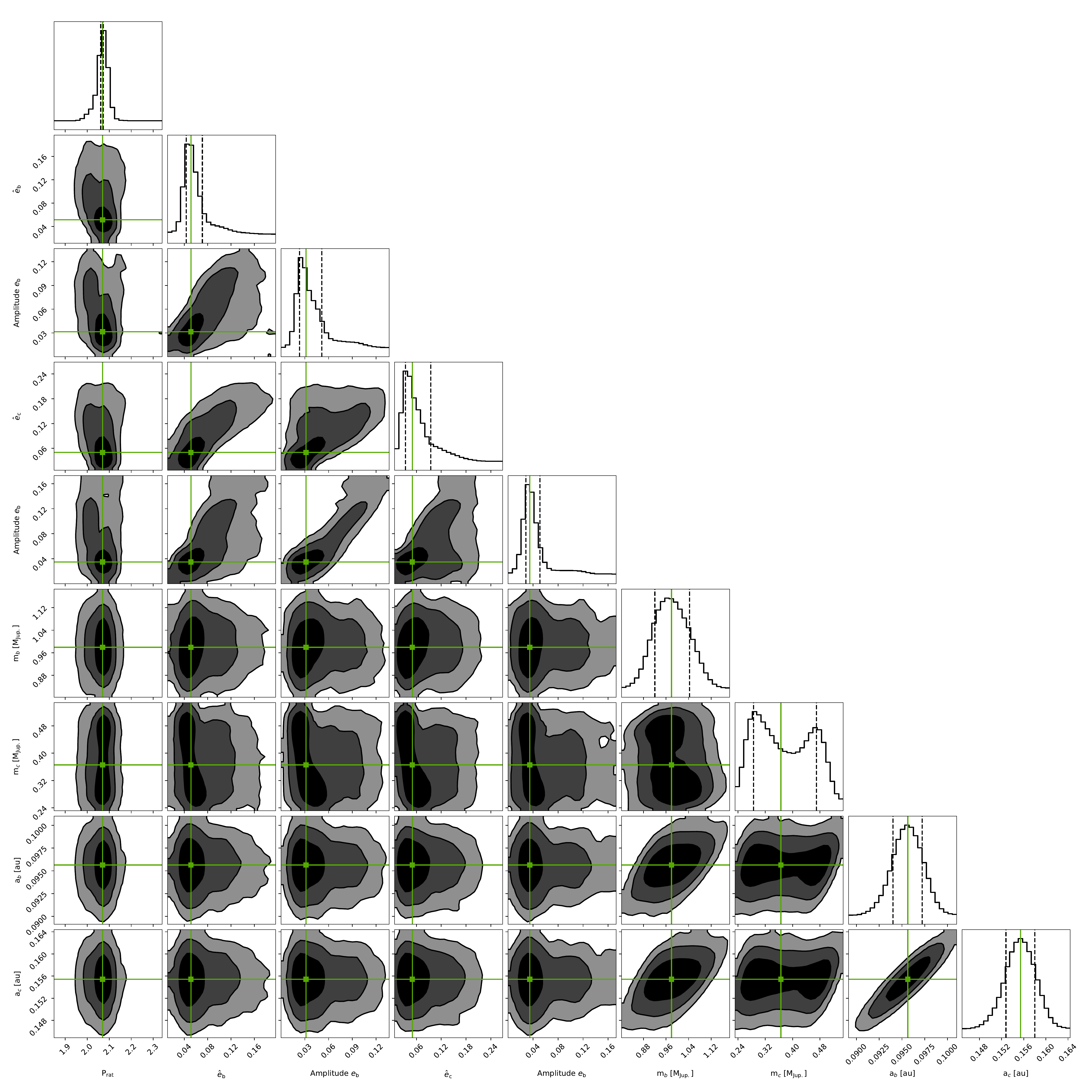} 

 \end{array} $
\end{center}

\caption{Posteriors of the dynamical properties at the 2:1 period ratio commensurability of the two-planet system TOI-2202 achieved by randomly drawing 10\,000 samples from the global posterior of the self-consistent dynamical model. 
Each sample is tested for stability and the overall dynamical properties at the 2:1 period ratio commensurability evaluated. The derived dynamical parameters are: mean period ratio $P_{\rm rat.}$, mean eccentricities $\hat e_{\rm b}$, $\hat e_{\rm c}$, their end-to-end amplitudes Ampl. $e_{\rm b}$, Ampl. $e_{\rm c}$, and their dynamical masses and semi-major axes. Note that the mass of TOI-2202 c is bimodal. The posteriors of the 2:1 MMR dynamical parameters  $\Delta\omega$, $\theta_1$,  and $\theta_2$ are not shown, since these exhibit circulation between 0 and 2$\pi$. 
The black contours on the 2D panels represent the 1, 2 and 3$\sigma$ confidence level of the overall posterior samples, whereas the green crosses indicate the median values of the derived posteriors.}
 
\label{dyn_samp} 
\end{figure*}

\begin{figure*}[tp]
\begin{center}$
\begin{array}{ccc}

\includegraphics[width=6cm]{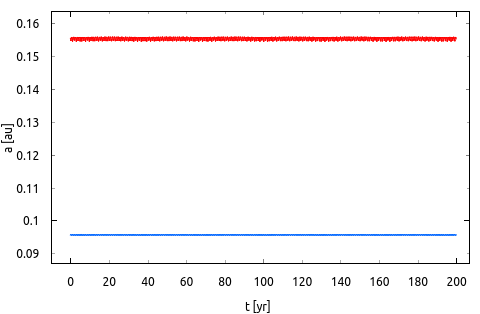} 
\includegraphics[width=6cm]{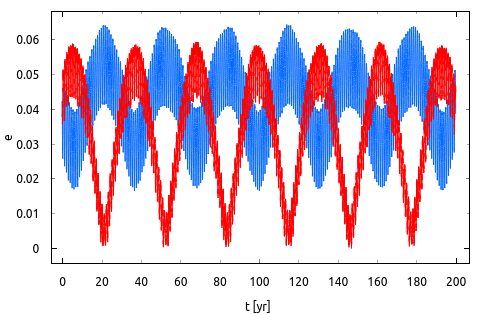} 
\includegraphics[width=6cm]{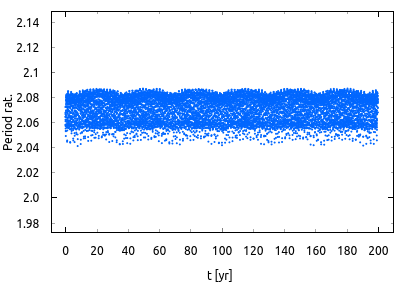}\\

\includegraphics[width=6cm]{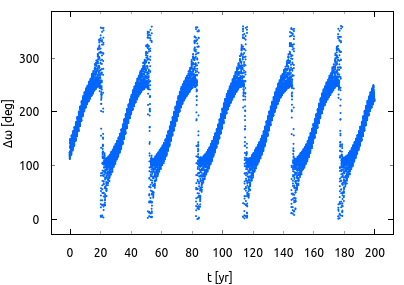} 
\includegraphics[width=6cm]{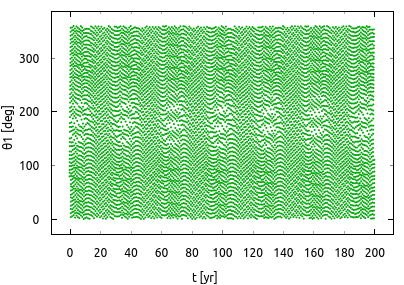}
\includegraphics[width=6cm]{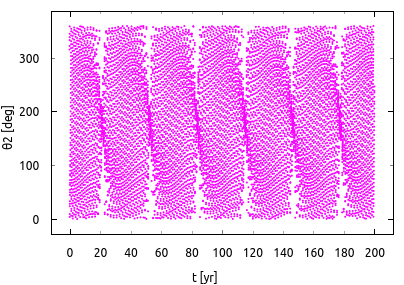}\\

\end{array} $
\end{center}

\caption{Orbital evolution of the TOI-2202 system for a 200 yr long N-body integration using Wisdom-Holman scheme.
    {\em Top panels}: Evolution of the planetary semi-major axes $a_{\rm b}$ and $a_{\rm c}$ (left), and the planetary eccentricities $e_{\rm b}$ and $e_{\rm c}$ (middle) of the best-fit N-body model, and the period ratio (right).
    {\em Bottom panels}: Evolution of the apsidal alignment angle $\Delta\varpi$ = $\varpi_{\rm b}$ - $\varpi_{\rm c}$ (left), 
    and the resonance angles $\theta_1$ and $\theta_2$ (middle, and right).
    Despite being close to the 2:1 MMR 
    the pair seems to reside outside of the low-order 2:1 MMR, as no libration in any of the 
    resonance angles is observed and the mean period evolution is osculating above the 2:1 period ratio.\looseness=-5
}
 
\label{evol_plot} 
\end{figure*}

Similarly to \autoref{Sec4.4}, we performed a 
nested sampling test, which we use to estimate the fitted parameter posteriors and confidence intervals, with a numerical time step in the dynamical model set to 0.1 days. 
We fitted the $\tess$ and the CHAT light curves adopting central body mass of 0.823 $M_\odot$, 
and free orbital parameters, which for the two planets are the planetary orbital period $P$, eccentricity $e$ and argument of periastron $\omega$,
and orbital inclination $i$. The dynamical modeling scheme within the {\tt Exo-Striker}
fits the osculating orbital parameters for a given epoch (i.e., non-static, perturbed orbital parameters), and thus, it requires the mean-anomaly $M_0$ parameterization, instead of the time of inferior transit conjunction $t_{0}$, which is commonly used in Keplerian models when fitting transit light curves. For each sampling iteration, the {\tt Exo-Striker} computes the perturbed orbital elements and transit times, which allows the precise modeling of the light curves. Since only TOI-2202 b transits, we only fit the light curves with the planetary scaled semi-major axis, $a_b/R_\star$, and radius $R_b/R_\star$, whereas for TOI-2202 c these are unconstrained. 
However, we 
allowed $i_c$ to vary in the fitting to account for mutual orbital inclinations, and therefore non-co-planar orbital dynamics.
Thus, we allowed TOI-2202 c to transit in our orbital analysis, although such models are naturally penalized by a poorer $-\ln\mathcal{L}$. 
The fitted orbital parameters for the RV model are shared with those of the transit model, while the RV signal semi-amplitude $K$ parameters for each planet constrain the planetary masses in the dynamical model. For the $\tess$ and the CHAT light curves, we adopted different quadratic limb-darkening models and varied the quadratic limb-darkening parameters $u_1$ and $u_2$ for each instrument.
We also varied the flux offset and jitter parameter of each transit light curve, and the offset and jitter parameters of each RV data set.

\begin{figure}[tp]
\begin{center}$
\begin{array}{ccc}

\includegraphics[width=4.4cm]{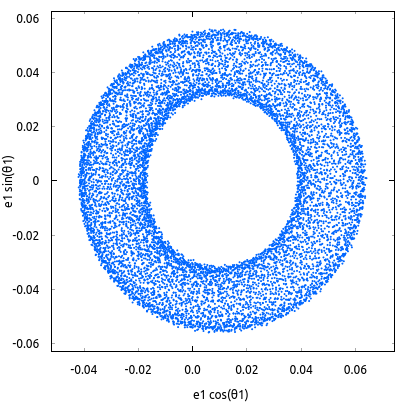}
\includegraphics[width=4.4cm]{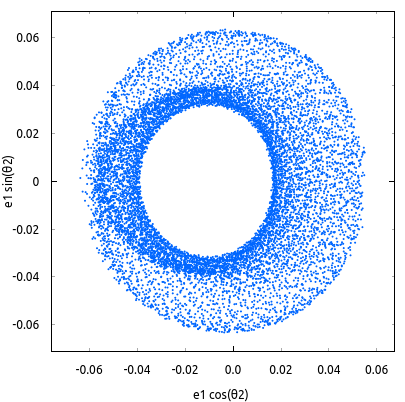}\\
\includegraphics[width=4.4cm]{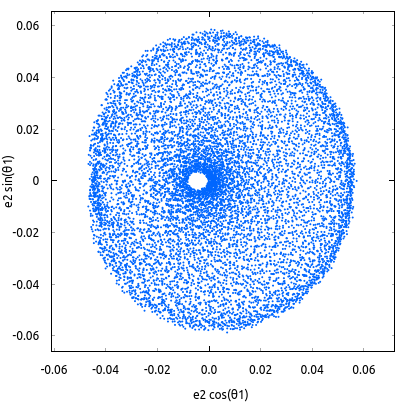}
\includegraphics[width=4.4cm]{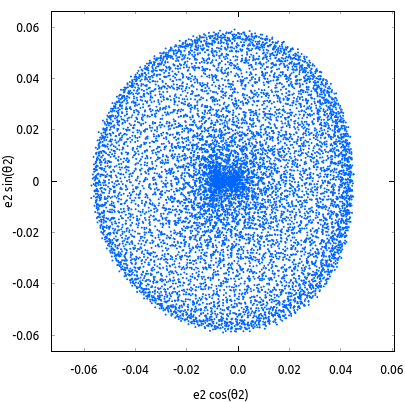}

\end{array} $
\end{center}

\caption{Same as in \autoref{evol_plot}, but represented in trajectory evolution of different combinations of $e_b$, $e_c$ and $sine$ and $cosine$ of the 2:1 MMR resonance angles $\theta_1$ and $\theta_2$. There is no observed libration around a fixed point in the trajectory evolution that could suggest a low-order mean-motion resonance in the 2:1 period ratio commensurability. 
}
 
\label{evol_plot2} 
\end{figure}

\autoref{tra_data} shows the transit component of the photo-dynamical model 
constructed together with a GP model for $\tess$, linear models for CHAT, and constrained by the RV data. \autoref{RV_data_GP} shows the RV component of the photo-dynamical model fitted to the FEROS, the HARPS, and the PFS data, including the GP model component.
The final estimates we derived from the joint photo-dynamical model posterior probability distribution are planetary periods of P$_b$ = 11.9101$_{-0.0036}^{+0.0022}$ days, and P$_c$ = 24.674$_{-0.034}^{+0.026}$ days, 
eccentricities of  $e_b$ = 0.042$_{-0.008}^{+0.025}$ and $e_c$ = 0.062$_{-0.021}^{+0.026}$, and dynamical masses of
$m_b$ = 0.978$_{-0.059}^{+0.063}$ $M_{\rm jup}$ and $m_c$ = 0.369$_{-0.084}^{+0.103}$ $M_{\rm jup}$. The mutual inclination, we constraint to $\Delta i$ = 6.56$_{-2.10}^{+1.92}$ deg.
We note that the posterior distribution for the mass TOI-2202 c planet is bimodal, with peaks at $\sim$ 0.30 $M_{\rm jup}$ and $\sim$ 0.47 $M_{\rm jup}$. We took the median of this bimodal distribution, meaning that it is possible that the mass uncertainties of TOI-2202 c are slightly underestimated. Also, it shows that a full photo-dynamical model can also suffer form mass-eccentricity ambiguities.
The full set of fitted and derived the orbital and physical planetary parameter estimates of TOI-2202\,b and c are listed in \autoref{NS_params}. The nuisance parameter estimates from the photo-dynamical nested sampling analysis are listed in \autoref{NS_params2}. The {\tt Exo-Striker} session, which contains the photo-dynamical model, priors, posterior distribution of all 52 fitted parameters, and the final correlation plot are available in \url{https://github.com/3fon3fonov/TOI-2202}.

We inspected the  photo-dynamical model residuals, and we did not detect additional periodic signals in the RVs or transit light curve. Thus, we concluded that the available data show evidence for only two planets in orbit around TOI-2202.

\begin{figure*} 
\begin{center}$
\begin{array}{ccc}

\includegraphics[width=8.9cm]{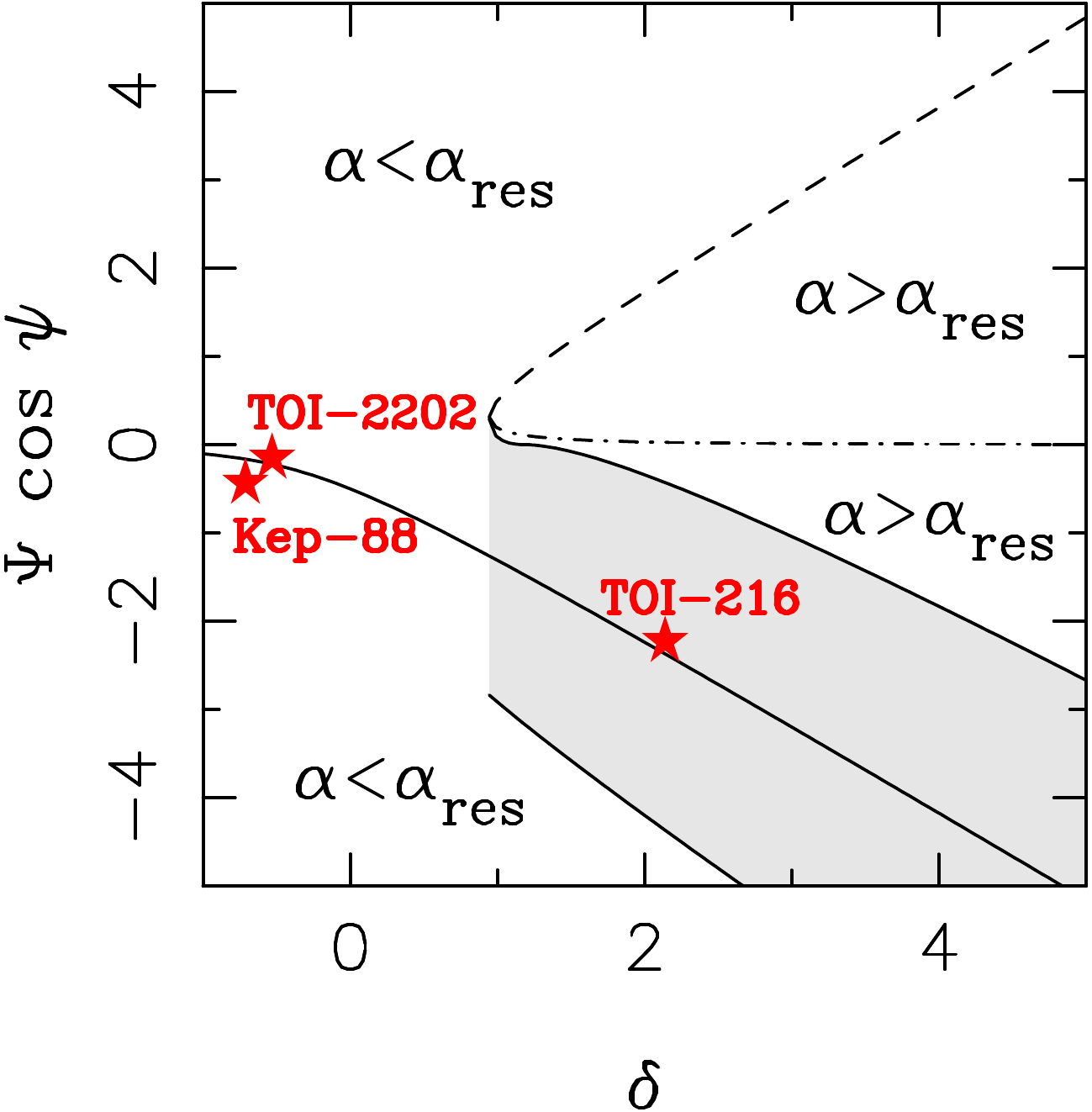}
\includegraphics[width=8.9cm]{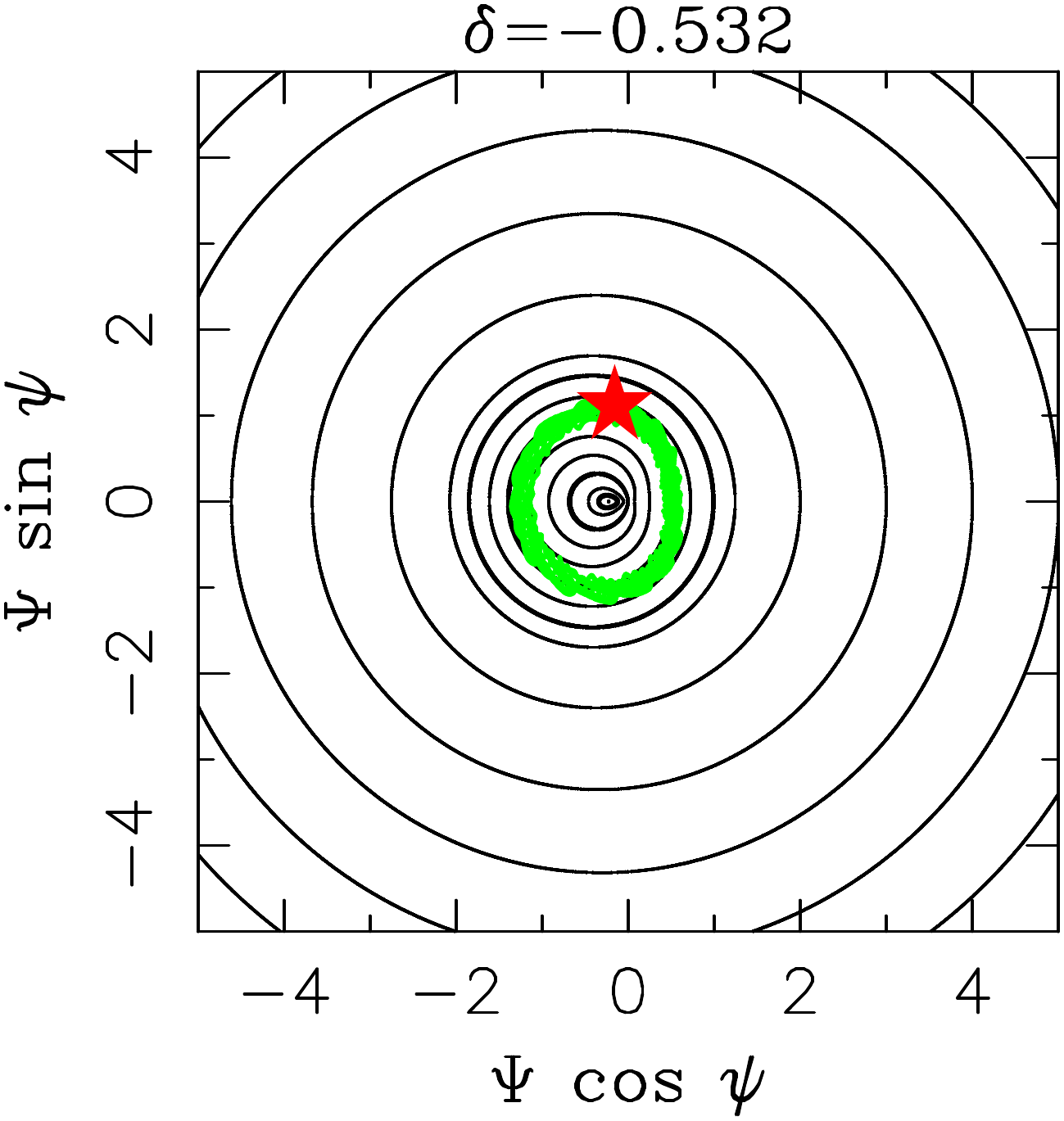}\\
 
\end{array} $
\end{center}

\caption{Left panel shows the 2:1 MMR structure diagram following \citet{Nesvorny2016}. Three planetary systems are plotted: TOI-2202 (this work), TOI-216 
\citep{Dawson2021} and Kepler-88 \citep{Nesvorny2013}. 
See \citet{Nesvorny2016} for definition
of parameters $\delta$ and $\Psi$, and the resonant angle $\psi$. 
Systems with $\alpha=a_{\rm b}/a_{\rm c}<\alpha_{\rm res}$ 
($\alpha>\alpha_{\rm res}$), where $\alpha=0.630$ corresponds to the exact resonance, have orbits just wide (narrow) of the resonance. The resonance region where $\psi$ librates is shaded. 
The separatrices and stable point are solid. The dotted line is an approximation of the stable point inside the resonance, as explained in  \citet{Nesvorny2016}. The dashed line is the unstable point. Dot-dashed is the stable point. See \citet{Nesvorny2016} for more explanations. 
Both TOI-2202 and Kepler-88 are wide of the resonance where $\psi$ circulates. TOI-216 is a resonant system. Right panel shows the trajectory N-body orbital evolution of the best photo-dynamical fit of TOI-2202, mapped onto resonant variables (green dots). The black curves are the analytic approximation of the resonant trajectories from \citep{Nesvorny2016}. There is a good correspondence between the numerical and analytical
evolution. The small differences between the two arise due to the neglected terms in the analytic expansion.\looseness=-5
}
 
\label{evol_plot3} 
\end{figure*}

\section{Dynamics and long-term stability}
\label{sec5}

\subsection{Numerical simulations}
\label{sec5.1}

We performed long-term stability and dynamical analysis of the TOI-2202 system using a custom version of the Wisdom-Holman N-body algorithm \citep[][]{Wisdom1991}, which directly adopts and integrates the Jacobi orbital elements from the {\tt Exo-Striker}.  Because of the relatively short orbital periods of the TOI-2202 planets, we chose a very small integration time step equal to 0.02\,d, which was necessary for the accurate numerical calculation analysis of the dynamical properties of the system. However, this short time step leads to severe numerical overhead, which limited our ability to test the system over the system's estimated age. For example, an N-body integration of the TOI-2202 system for only 10\,Myr takes approximately one day on a modern CPU. An educated guess, however, suggests that longer integration times are likely not necessary. The {\tt Exo-Striker} provides an instant Angular Momentum Deficiency \citep[AMD,][]{Laskar2017} monitoring, which indicates that given the estimated small orbital eccentricities,  semi-major axes, mutual orbital inclination, and planetary masses, the TOI-2202 system is AMD-stable. For more details on the calculation of the AMD-stability criteria, we refer to \citet[][]{Laskar2017}. Additionally, given the estimated semi-major axes and masses in the system, we can calculate the mutual Hill radii of the planets as:

\begin{equation}
 R_{\rm Hill,m}\approx \sqrt[3]{\frac{(m_b + m_c)}{3M_{\star}}}   \frac{(a_b + a_c)}{2}  \approx 0.01~{\rm au}
\label{eq:hill1}
\end{equation}



\noindent
and since a$_c$ -- a$_b$ = 0.06 au, then the planets are about 6 R$_{\rm Hill,m}$, which is above the $\sim$ 3.5 R$_{\rm Hill,m}$  threshold needed for the system to be considered Hill-stable  \citep[see,][]{Gladman1993}. In terms of AMD and Hill-stability, the TOI-2202  planetary system must be generally stable despite the close planetary orbits. Nevertheless, the AMD and Hill stability criteria do not account for the system's dynamics near mean motion resonances, thus can only be used as a proxy for long-term stability. Therefore, as a compromise, our long-term stability simulations of TOI-2202 system were performed for a maximum of 1\,Myr by numerically integrating 10\,000 randomly chosen samples of the achieved parameter posteriors from the photo-dynamical modeling scheme.   We ran our stability test in a modern 40-CPU {\tt Intel Xeon} based workstation, which took about four weeks to complete, and which we find to be reasonable for the N-body dynamical analysis in this work.

For each posteriors sample integration, we automatically monitored the evolution of the planetary semi-major axes and eccentricities as a function of time to assure that the system remains regular and well separated at any given time of the simulation. Any deviation of the planetary semi-major axes by more than 20\% from their starting values, or eccentricities leading to crossing orbits, were considered unstable.\looseness=-2

Given the period ratio of the system close to the 2:1 commensurability, we inspected the first-order MMRs angles $\theta_1$ and $\theta_2$ , which are defined as:
\begin{equation}
\theta_1=\lambda_b-2\lambda_c+\varpi_b, \
\theta_2=\lambda_b-2\lambda_c+\varpi_c, 
\end{equation}

\noindent
where $\varpi_{\rm b,c}=\Omega_{\rm b,c}+\omega_{\rm b,c}$ are the planetary longitudes of periatron and $\lambda_{\rm b,c}$=M$_{0 \rm b,c}+\varpi_{\rm b,c}$ are the mean longitudes, respectively.
We also monitored for libration of the secular apsidal angle $\Delta\omega$, which is defined as:
\begin{equation}
\Delta\omega=\theta_1-\theta_2=\varpi_b-\varpi_c, 
\end{equation}
which indicates if the dynamics of the system is dominated by secular interactions, exhibiting
apsidal libration in alignment ($\Delta\omega$ librating around 0$^{\circ}$), anti-alignment ($\Delta\omega$ librating around 180$^{\circ}$), or an asymmetric libration.

The results from our long-term stability analysis indicate that all examined 10,000 samples are stable for 1\,Myr with very similar dynamical behavior. \autoref{dyn_samp} shows the derived posteriors of the dynamical properties of the studied 10,000 samples. 
The distribution of dynamical parameters reveals low-eccentricity evolution,
but despite being close to the 2:1 MMR, the TOI-2202 pair seems to reside outside of the low-order eccentricity type 2:1 MMR.
The mean period ratio evolution is osculating around 2.07, 
while
we did not detect libration of the resonance angles  $\theta_1$, $\theta_2$, and the apsidal alignment angle $\Delta\omega$. The posterior distributions of $\theta_1$, $\theta_2$, and $\Delta\omega$ are consistent with circulation, with libration amplitudes between 0$^\circ$ and 360$^\circ$.

\autoref{evol_plot} shows a 200 yr extent of the dynamical evolution of the photo-dynamical fit with a maximum $-\ln\mathcal{L}$ from the posteriors probability samples (see \autoref{NS_params}), which is representative of the overall dynamics of the tested posterior samples. \autoref{evol_plot} shows the evolution of the semi-major axes $a_{\rm b}$ and $a_{\rm c}$, eccentricities $e_{\rm b}$ and $e_{\rm c}$, the period ratio, the apsidal alignment angle $\Delta\varpi$, and the characteristic 2:1 MMR angles $\theta_1$, and $\theta_2$, respectively. No resonance angle libration is observed. For the same fit, \autoref{evol_plot2} shows the trajectory evolution of different combinations of $e_b$, $e_c$ and $sine$ and $cosine$ functions of the 2:1 MMR resonance angles $\theta_1$ and $\theta_2$. There is no observed libration around a fixed point in the trajectory evolution that could suggest a 2:1 MMR. Similar trajectory evolution is observed in the majority of the posterior samples that are within the 1 $\sigma$ credible interval.\looseness=-5


\subsection{Analytical analysis}
\label{sec5.2}

The eccentricities of both planets are low and when that is the case, the resonant and near-resonant 
dynamics can be studied analytically following \citet{Nesvorny2016}. There are three variables 
to consider, each of them being a combination of orbital elements. Constant $\delta$ is an orbital 
invariant that defines the position of the system relative to the 2:1  commensurability, 
the resonant angle $\psi$ is a combination of $\theta_1$ and $\theta_2$ and 
variable $\Psi$ is a combination of planetary masses, semi-major axes 
and eccentricities \citep[see,][for details]{Nesvorny2016}. 
\autoref{evol_plot3} shows position of TOI-2202 in the context of the dynamical variables $\delta$,  $\psi$, and  $\Psi$. The resonant librations of 
$\psi$ can only happen for $\delta>0.945$. The best-fit and median values of TOI-2202, listed in \autoref{NS_params}, lead to $\delta \simeq - 0.532$, and $- 0.77$, respectively, and the system is therefore firmly 
outside the libration region. For TOI-2202, $\Psi$ and $\psi$ follow a deformed circle that is slightly offset 
from the origin, which means that $\psi$ circulates and $\Psi$ 
oscillates ($0.4\lesssim \Psi \lesssim 1.1$).

Analytic expressions can be used to relate the TTVs measured for TOI-2202 to the system's architecture. Given 
that the two orbits are {\it not} librating in the 2:1 resonance, here we use analytic TTV expressions given 
in \citep{Agol2016}. Eq. (6) in that paper gives the non-resonant TTVs to the first order in orbital
eccentricities. We only consider the highest amplitude term for the inner planet:
\begin{equation}
\delta t_1 = {P_b \over 2 \pi} {m_c \over M_*} f^{(-2)}_{1,1} e_c\sin \theta_2 \ ,   
\end{equation} 
where $f^{(-2)}_{1,1} \simeq 0.05$. The expected amplitude
of TTVs is therefore $A_{\rm TTV} \simeq 0.05$ days, which compares well with the measured amplitude (see, \autoref{TTV_plot1}).
Other terms, including TTV chopping during planet conjunctions \citep{Nesvorny2014, 
Deck2015} are responsible for the slight deviations of measured TTVs from a perfect sinusoid. 
These terms are important to break degeneracies in the TTV inversion.     
The expected TTV period is the period of the $\theta_2$ angle. Neglecting the long-term evolution of $\varpi_c$, 
the TTV period can be approximated by the super-period:
\begin{equation}
P_{\rm TTV}=\left[ {1 \over P_b} - {2 \over P_c} \right ]^{-1}.  
\end{equation}

\noindent
Adopting the best-fit orbital periods from \autoref{NS_params}, we obtain $P_{\rm TTV} \simeq 340$ days, which is a good match to \autoref{TTV_plot1}.

\section{Summary and Conclusions}
\label{sec6}

We report the discovery of a compact Jovian-mass pair of planets around the K-dwarf star TOI-2202, for which we estimate a stellar mass of 0.823$_{-0.023}^{+0.027}$ M$_\odot$ and a radius of 0.794$_{-0.007}^{+0.007}$ R$_\odot$. This discovery was possible thanks to $\tess$, which revealed the transiting warm Jovian mass planet TOI-2202\,b that transits with a period of about 11.91 days.  The ten $\tess$ transits of TOI-2202\,b detected on the FFI, together with three follow-up light curves obtained with the CHAT robotic telescope, show strong TTVs with an amplitude of about 1.2 hours, suggesting the presence of a second, non-transiting, massive body that perturbs the transiting planet.

A precise Doppler spectroscopy follow-up with FEROS, HARPS, and PFS firmly confirmed the transiting candidate's planetary nature, pointing to a Jovian-mass planet with a dynamical mass of m$_{\rm p}\sim$1.0 M$_{\rm Jup}$. We performed an extensive analysis of the RVs and transit data, and we reveal an outer Saturn-mass companion with a mass of m$_{\rm p}\sim$0.4 M$_{\rm Jup}$ and a period of 24.67 days, which puts the warm pair of massive planets near the 2:1 period ratio commensurability. 
The mass and period of the outer planet, TOI-2202 c, were indirectly revealed thanks to the dynamical orbital analysis since the available transit and Doppler data did not directly support its presence.   
From our combined Doppler and photo-dynamical modeling scheme, we obtain a semi-amplitude $K_c$ =  29.3$_{-6.6}^{+8.3}$ m\,s$^{-1}$ of TOI-2202 c, which is significant, and given the precision of the RV data, and it should have been detected by our MLP period search (see the one-planet RV residuals MLP in   \autoref{MLP_results}).
We attribute the non-detection of the RV signal induced by TOI-2202\,c to the combination of at least four important effects: (i) All RV datasets have a notable white-noise in terms of ``RV" jitter (see \autoref{NS_params2}). The larger variance of the data, contributes in reducing the MLP power.   
(ii) The stellar activity, which seems to strongly influence the $\tess$ light curve and the RV data. 
We retain a strong quasi periodic signal with a period of (transit+RV) GP$_{\rm Rot.}$ = 24.1$_{-1.8}^{+2.3}$\,d, and an RV amplitude of GP$_{\rm Rot.}$ Amp. = 585.3$_{-244.3}^{+229.1}$ m$^{2}$\,s$^{-2}$ ($\sim$ 24.2$_{-16.6}^{+15.1}$ m\,s$^{-1}$), which are inconveniently similar to those of TOI-2202\,c (see e.g., bottom panel of \autoref{RV_data_GP}). The TOI-2202 case amply demonstrates the modeling difficulties which arise when stellar activity and Doppler planetary signals are nested within similar periods.
(iii) In Doppler observations, a pair of planets in
low-eccentricity orbits near a 2:1 MMR can be misinterpreted as a single planet \citep{Escude2010,Wittenmyer2013,Kuerster2015,Boisvert2018,Hara2019}. Thus, it is possible that some of the signal from TOI-2202 c may have been fit by a change in the estimated eccentricity of TOI-2202 b.
(iv) Last, but not least, the planets' strong mutual perturbations also impact the MLP period search.
The temporal baseline of our RV data is only $\sim$ 1\,yr, within which however, our N-body simulations reveal strong end-to-end secular variations of e.g., 2.042 $< P_{\rm rat.} < $ 2.085, and 110$^{\circ}$ $< \Delta\omega <$ 150$^{\circ}$,  which result in RV-signal phase shifts, and period osculation (as well in TTVs). 
All these effects contribute to a certain degree in blurring the Doppler-induced planetary signals. While TOI-2202 b is massive enough to induce a notable RV signal, the Doppler signal of the less massive outer planet TOI-2202 c seems to be buried in noise.

Finally, additional non-transiting planets could reside in the system, but we cannot disentangle their complex contribution to the observed RV signal. While such a possibility is valid, we do not have solid evidence that this could be the case. We note that while TOI-2202\,b and TOI-2202\,c orbit far enough from one another to be stable as a two-planet system, systems with three or more planets require larger separations for stability \citep[e.g.,][]{Chambers1996,Pu2015,Petit2020,Lissauer2021}. Thus, if any other planets are present in this system, their orbits are likely to be more distant (in terms of Hill sphere radii) from the two known planets than these two planets are from one another.

 
Our numerical and analytical dynamical analysis of the system configuration revealed that the Jovian-mass pair is actually outside of the exact 2:1 MMR.
We ruled out a 2:1 MMR librating configurations of TOI-2202, based on the available transit and RV data. 
The osculating period ratio of TOI-2202 is a little above 2, which falls within the peak of the distribution of period ratios of planet pairs observed by Kepler \citep{Lissauer2011, Fabrycky2014}. However, the Kepler sample is dominated by super-Earths and sub-Neptunes, and not warm, giant planets like TOI-2202 b \& c, which have masses consistent with Jupiter and Saturn.

There are several possible scenarios that could explain the non-resonant orbital configuration of the TOI-2202 pair of massive planets; (i) The giant planets formed in-situ near resonance and did not experience sufficient migration or eccentricity damping to capture them in resonance \citep{Lee2014,Boley2016,Batygin2016}.
(ii) The giant planets migrated but did not get captured into resonance, maybe due to over-stable libration \citep{Goldreich2014}, turbulence in the disk, or perturbations from other planets.
(iii) The giant planets were once in resonance, but dislodged through a dynamical interaction with one or more undetected other planets in the system \citep{Ford2005,Raymond2009,Carrera2019}.

A few known systems similar to TOI-2202 exist. 
Apart from TOI-216 \citep{Dawson2019, Dawson2021}, another very similar system of a warm Jovian-Saturn-mass pair in a possible 2:1 MMR is HD\,27894 \citep{Trifonov2017}, which is a 0.8 M$_\odot$ K dwarf with an inner massive pair of planets with periods of $\sim$18 d and $\sim$36 d.
The outer planet in the  HD\,27894  system is $\sim$4 times less massive than the inner, analogous to TOI-216 and TOI-2202. Then is the TOI-2525  system (Reichardt in prep.), which shows a very similar physical configuration as TOI-2202. The star TOI-2525 is a K dwarf similar to TOI-2202, which is orbited by a massive pair of inner Jovian planet and an external Saturn with a period ratio above 2:1  period ratio commensurability, but is also not resonant.
It seems plausible that such warm, near-resonant systems consistent with a Jovian-Saturn mass pair are found around K-dwarfs.

The TOI-2202 system is very intriguing because of the warm Jovian and Saturn-mass planet pair near the 2:1 MMR, which is a rare configuration, and their formation and dynamical evolution are still not well understood.  Thus, the TOI-2202 system is an essential addition to the
planetary occurrence statistics, and it could shed new light on planetary formation and evolution.

\acknowledgements

This research has made use of the Exoplanet Follow-up Observation Program website, which is operated by the California Institute of Technology, under contract with the National Aeronautics and Space Administration under the Exoplanet Exploration Program.
Funding for the $\tess$ mission is provided by NASA's Science Mission directorate.
This paper includes data collected by the $\tess$ mission, which are publicly available from the Mikulski Archive for Space Telescopes (MAST).
Resources supporting this work were provided by the NASA High-End Computing (HEC) Program through the NASA Advanced Supercomputing (NAS) Division at Ames Research Center for the production of the SPOC data products.
This work was supported by the DFG Research Unit FOR2544 ``Blue Planets around Red Stars", project No. RE 2694/4-1.
M.H.L. was supported in part by Hong Kong RGC grant HKU 17305618.
AJ, RB, MH and FR acknowledge support from ANID -- Millennium  Science  Initiative -- ICN12\_009. AJ acknowledges additional support from FONDECYT project 1210718.
MRD acknowledges the support by CONICYT-PFCHA/Doctorado Nacional-21140646, Chile.
TD acknowledges support from MIT's Kavli Institute as a Kavli postdoctoral fellow.
JIV acknowledges support of CONICYT-PFCHA/Doctorado Nacional-21191829.
D. D. acknowledges support from the TESS Guest Investigator Program grant 80NSSC19K1727 and NASA Exoplanet Research Program grant 18-2XRP18 2-0136.
 We thank the anonymous reviewer for very helpful comments and suggestions.
\facilities{$\tess$, CHAT-0.7m, MPG-2.2m/FEROS, ESO-3.6m/HARPS, Magellan-6.5m/PFS}

\software{
          Exo-Striker~\citep{Trifonov2019_es},
          CERES~\citep{ceres},
          ZASPE~\citep{zaspe},
          tesseract~(Rojas, in prep.),
          TESSCut~\citep{TESSCut},
          lightkurve~\citep{lightkurve},
          emcee~\citep{emcee},
          corner.py~\citep{corner},
		  dynesty~\citep{Speagle2020},
          batman~\citep{Kreidberg2015},
          celerite~\citep{celerite},
          wotan~\citep{Hippke2019},
           transitleastsquares~\citep{Hippke2019b}
          }

\clearpage

\bibliographystyle{aasjournal}
\bibliography{bibliography}

\appendix

 \setcounter{table}{0}
\renewcommand{\thetable}{A\arabic{table}}

\setcounter{figure}{0}
\renewcommand{\thefigure}{A\arabic{figure}}

\begin{figure*}[tp]
\begin{center}$
\begin{array}{ccc}

\includegraphics[width=18cm]{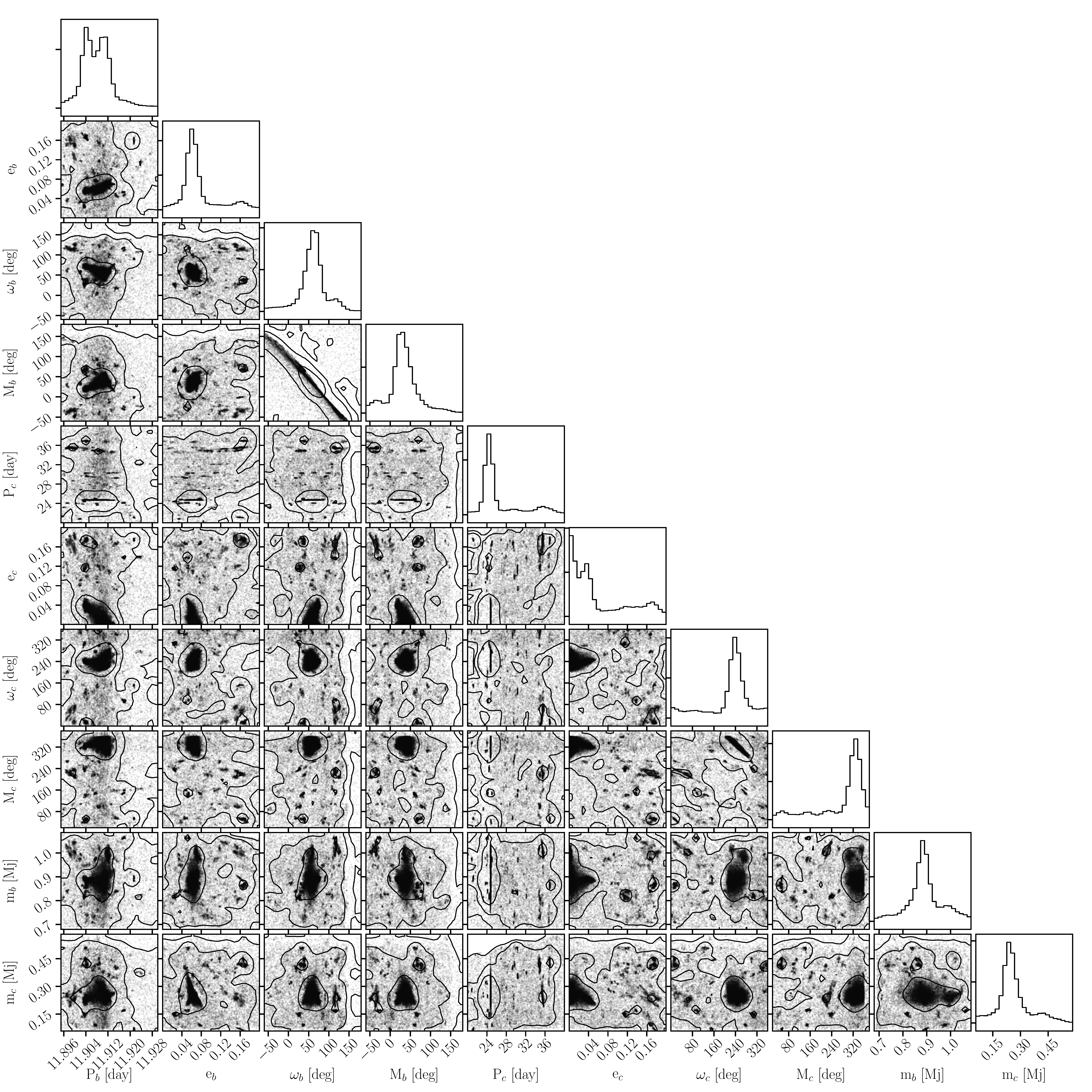} 
 \end{array} $
\end{center}

\caption{{\tt Exo-Striker} global parameters search results of the $\tess$ and CHAT TTVs of TOI-2202\,b 
done with a Nested sampling scheme assuming a coplanar, edge-on and prograde 
two-planet system fitted with a self-consistent dynamical model. 
The black contours on the 2D panels represent the 1, 2, and 3$\sigma$ confidence level of the overall nested sampling samples. 
The distribution of orbital parameters reveals that the observed TTVs of TOI-2202\,b is multi-modal, but is 
most likely induced by an exterior Saturn mass planet close whose orbital period 
is close to the first-order eccentricity-type 2:1 MMR with TOI-2202\,b.
}
 
\label{Nest_samp_ttv} 
\end{figure*}

\begin{figure*}[tp]
\begin{center}$
\begin{array}{ccc}

\includegraphics[width=18cm]{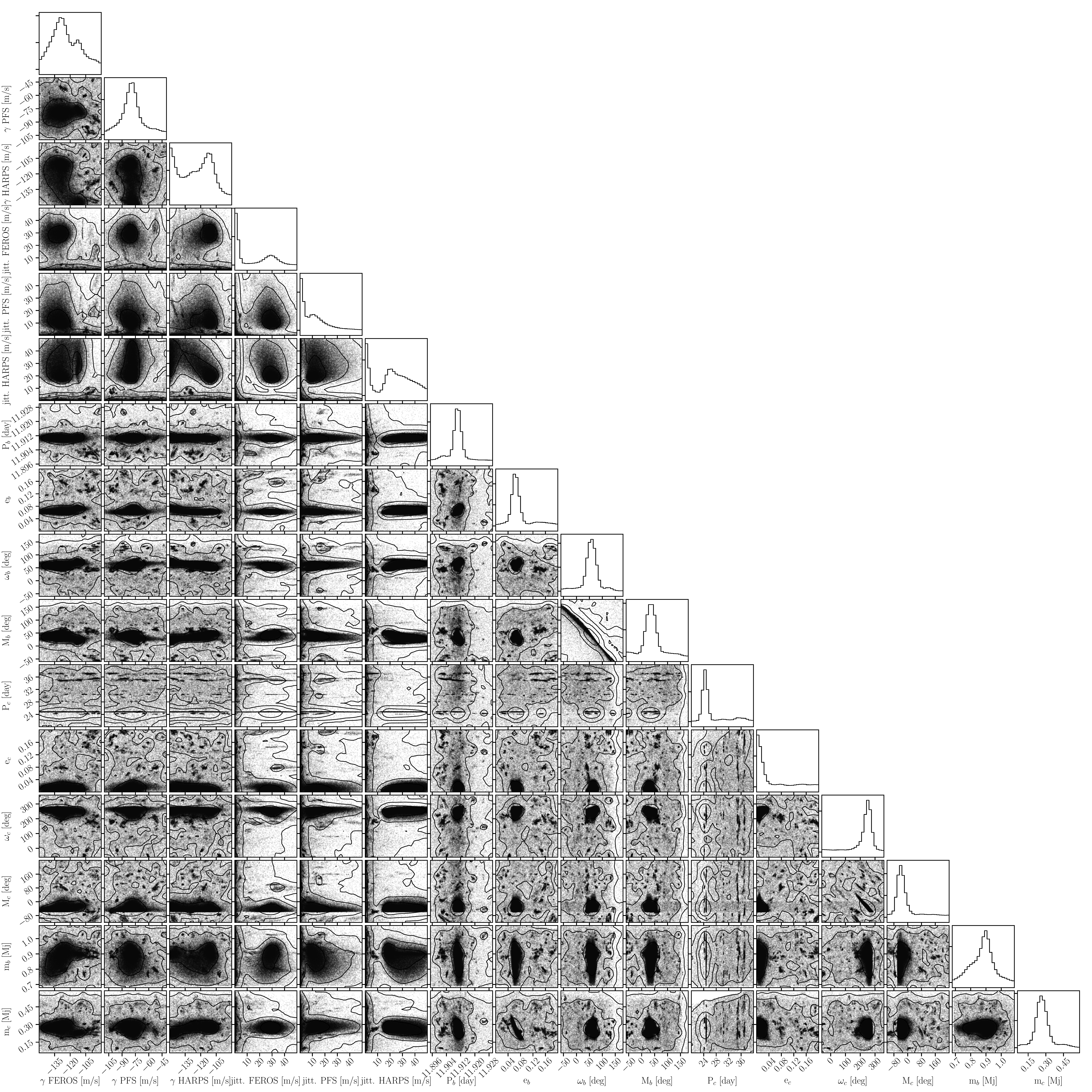} 
 \end{array} $
\end{center}

\caption{Same as \autoref{Nest_samp_ttv} but including the Doppler data in the 
global nested sampling scheme. The nested sampling distribution shows the distribution of orbital parameters 
consistent with the FEROS, PFS, and HARPS RV data and the $\tess$ and CHAT TTVs of TOI-2202\,b  
assuming a coplanar, edge-on and prograde two-planet system fitted with a self-consistent dynamical model.
The black contours on the 2D panels represent the 1, 2, and 3$\sigma$ confidence level of the overall nested sampling samples. 
}
 
\label{Nest_samp_ttv2} 
\end{figure*}

%
%
%
%
%
%
%
%
%
%
%

\begin{table}
\caption{FEROS Doppler measurements of TOI-2202  } 
\label{table:TIC358107516_FEROS} 

\centering  

\begin{tabular}{c c c } 

\hline\hline    
\noalign{\vskip 0.5mm}

Epoch [JD] & RV [m\,s$^{-1}$] & $\sigma_{RV}$ [m\,s$^{-1}$]   \\  

\hline     
\noalign{\vskip 0.5mm}    

2458542.6107   &   $-$201.2   &    10.7      \\ 
2458544.5814   &   $-$210.0   &    11.0      \\ 
2458546.5567   &   $-$164.6   &    12.5      \\ 
2458549.5470   &   $-$27.8   &    10.5      \\ 
2458550.5787   &   $-$68.1   &    10.3      \\ 
2458551.5303   &   $-$82.7   &    13.1      \\ 
2458553.5551   &   $-$175.7   &    20.2      \\ 
2458554.5343   &   $-$152.9   &    21.3      \\ 
2458555.5589   &   $-$200.5   &    11.3      \\ 
2458556.5614   &   $-$188.1   &    10.6      \\ 
2458557.5492   &   $-$203.7   &    13.5      \\ 
2458558.5407   &   $-$40.5   &    18.0      \\ 
2458559.5642   &   26.5   &    14.9      \\ 
2458566.5039   &   $-$224.3   &    10.0      \\ 
2458567.5359   &   $-$171.9   &    10.3      \\ 
2458568.5287   &   $-$235.4   &    10.3      \\ 
2458569.5235   &   $-$251.8   &    12.1      \\ 
2458570.5308   &   $-$214.1   &    15.2      \\ 
2458572.5180   &   $-$140.6   &    12.3      \\ 
2458653.8986   &   $-$72.0   &    14.6      \\ 
2458660.8934   &   $-$25.6   &    23.3      \\ 
2458676.9125   &   $-$154.9   &    15.6      \\ 
2458782.7537   &   $-$193.2   &    9.2      \\ 
2458793.5853   &   $-$205.9   &    9.0      \\ 
2458805.7142   &   $-$209.8   &    8.8      \\ 
  
\hline           
\end{tabular}

\end{table}

\begin{sidewaystable}
\resizebox{0.62\textheight}{!}
     {\begin{minipage}{1.1\textwidth}

\caption{HARPS Doppler measurements of TOI-2202  } 
 \label{table:TIC358107516_HARPS} 
\centering  


\begin{tabular}{ccccccccccccccccccccccccccccccccccc} 

\hline\hline    
\noalign{\vskip 0.5mm}
Epoch [JD] & 
DRS RV [m\,s$^{-1}$]  & 
DRS $\sigma_{RV}$ [m\,s$^{-1}$]   & 
BIS [m\,s$^{-1}$] & 
$\sigma_{BIS}$ [m\,s$^{-1}$]   & 
 Contrast &
$\sigma_{Contrast}$  & 
FWHM [m\,s$^{-1}$] & 
$\sigma_{FWHM}$ [m\,s$^{-1}$]& 
SERVAL RV [m\,s$^{-1}$] &
SERVAL $\sigma_{RV}$ [m\,s$^{-1}$]   & 
CRX [m\,s$^{-1}$] & 
$\sigma_{CRX}$ [m\,s$^{-1}$] &
 dLW [m\,s$^{-1}$] & 
$\sigma_{dLW}$  &
H$_{\alpha}$& 
 $\sigma_{H_{\alpha}}$  &
NaD I& 
$\sigma_{NaD I}$ &
NaD II & 
$\sigma_{NaD II}$ \\  

\hline     
\noalign{\vskip 0.5mm}    
2458762.8145   &   -52.26    &    6.52 &   10.434    &    0.104  &   56.009   &    0.560 &   7.255   &    0.073 &  54.78  & 3.31 &    -12.373   &    27.905 &   0.553    &    1.638 &   0.525   &    0.005 &   0.214   &    0.006 &   0.286   &    0.008      \\ 
2458764.7474   &   -8.08     &    6.55 &   -35.737   &    -0.357 &   55.505   &    0.555 &   7.273   &    0.073 & 129.14  & 2.30 &     6.715    &    16.390 &   -0.133   &    1.152 &   0.526   &    0.004 &   0.263   &    0.005 &   0.309   &    0.006      \\ 
2458773.7777   &   -149.33   &    8.32 &   -6.315    &    -0.063 &   53.973   &    0.540 &   7.386   &    0.074 &  -81.35 & 4.07 &    80.253    &    46.840 &   5.546    &    1.836 &   0.525   &    0.006 &   0.280   &    0.008 &   0.335   &    0.010      \\ 
2458774.7145   &   -79.47    &    7.06 &   4.569     &    0.046  &   55.332   &    0.553 &   7.209   &    0.072 &  185.53 & 3.01 &    -105.263  &    41.788 &   9.357    &    1.488 &   0.515   &    0.005 &   0.233   &    0.006 &   0.271   &    0.007      \\ 
2458777.6782   &   -36.10    &    6.68 &   -27.291   &    -0.273 &   56.648   &    0.566 &   7.241   &    0.072 &  88.47  & 4.07 &    42.774    &    33.940 &   0.901    &    1.807 &   0.536   &    0.007 &   0.293   &    0.009 &   0.334   &    0.010      \\ 
2458807.7079   &   -186.62   &    6.06 &   -30.652   &    -0.307 &   55.845   &    0.558 &   7.230   &    0.072 &  -44.15 & 2.61 &    7.185     &    20.158 &   -3.248   &    1.398 &   0.521   &    0.004 &   0.209   &    0.005 &   0.273   &    0.006      \\ 
2458811.7423   &   -29.35    &    6.56 &   -2.856    &    -0.029 &   56.013   &    0.560 &   7.193   &    0.072 &  94.87  & 3.40 &    9.363     &    26.927 &   -1.325   &    1.567 &   0.522   &    0.005 &   0.202   &    0.006 &   0.275   &    0.008      \\ 
2458813.7032   &   -40.23    &    6.41 &   10.524    &    0.105  &   56.207   &    0.562 &   7.229   &    0.072 &  69.06  & 3.40 &    -3.834    &    29.515 &   1.149    &    1.420 &   0.530   &    0.005 &   0.186   &    0.006 &   0.251   &    0.008      \\ 
2458831.6523   &   -177.98   &    6.45 &   -20.233   &    -0.202 &   55.180   &    0.552 &   7.269   &    0.073 & -44.42  & 2.29 &     3.715    &    15.006 &   -1.157   &    0.938 &   0.529   &    0.004 &   0.178   &    0.004 &   0.256   &    0.005      \\ 
2458834.7019   &   -52.83    &    6.73 &   -17.317   &    -0.173 &   55.794   &    0.558 &   7.272   &    0.073 &  75.55  & 3.15 &    32.535    &    24.575 &   -0.829   &    1.258 &   0.530   &    0.005 &   0.193   &    0.006 &   0.257   &    0.007      \\ 
2458838.6934   &   -94.04    &    8.57 &   37.192    &    0.372  &   55.380   &    0.554 &   7.338   &    0.073 &  -15.09 & 5.24 &     -37.526  &    39.004 &   4.921    &    2.431 &   0.532   &    0.008 &   0.173   &    0.010 &   0.248   &    0.012      \\ 
2458840.7025   &   -152.85   &    6.55 &   -21.212   &    -0.212 &   55.780   &    0.558 &   7.287   &    0.073 &  -22.39 & 2.86 &    -9.285    &    23.949 &   -0.613   &    1.105 &   0.533   &    0.005 &   0.202   &    0.005 &   0.246   &    0.006      \\ 
2458852.6085   &   -139.95   &    6.32 &   -7.635    &    -0.076 &   55.724   &    0.557 &   7.247   &    0.072 &  -6.82  & 2.08 &    -39.069   &    15.335 &   -6.968   &    1.115 &   0.530   &    0.004 &   0.187   &    0.004 &   0.256   &    0.005      \\ 
2458869.5997   &   -95.12    &    7.62 &   7.111     &    0.071  &   54.063   &    0.541 &   7.293   &    0.073 &  63.63  & 2.53 &    -16.876   &    26.979 &   28.673   &    1.856 &   0.524   &    0.004 &   0.198   &    0.005 &   0.272   &    0.006      \\ 
2458883.5666   &   23.19     &    6.31 &   -6.497    &    -0.065 &   55.299   &    0.553 &   7.236   &    0.072 &  144.58 & 3.10 &    5.812     &    23.744 &   -2.980   &    1.310 &   0.522   &    0.005 &   0.175   &    0.006 &   0.266   &    0.007      \\ 
2458883.5873   &   15.10     &    8.00 &   -0.441    &    -0.004 &   54.625   &    0.546 &   7.329   &    0.073 &  149.58 & 3.39 &    -29.540   &    26.864 &   -1.722   &    1.727 &   0.534   &    0.005 &   0.175   &    0.006 &   0.254   &    0.008      \\ 
2458886.5552   &   -76.03    &    6.81 &   13.288    &    0.133  &   55.998   &    0.560 &   7.253   &    0.073 &  60.78  & 3.65 &    -74.472   &    27.175 &   0.324    &    1.652 &   0.532   &    0.006 &   0.185   &    0.007 &   0.266   &    0.008      \\ 
2458890.5358   &   -218.06   &    7.52 &   -6.051    &    -0.061 &   54.543   &    0.545 &   7.352   &    0.074 &  -63.98 & 2.50 &    -29.871   &    23.701 &   -3.326   &    1.286 &   0.534   &    0.004 &   0.187   &    0.005 &   0.253   &    0.006      \\ 
2458898.5693   &   -130.24   &    8.32 &   23.765    &    0.238  &   55.213   &    0.552 &   7.312   &    0.073 &  22.72  & 3.99 &    -14.593   &    29.519 &   -0.539   &    1.581 &   0.518   &    0.006 &   0.196   &    0.008 &   0.254   &    0.010      \\ 
2458900.5619   &   -189.56   &    8.40 &   -35.994   &    -0.360 &   55.953   &    0.560 &   7.259   &    0.073 &  -32.42 & 4.35 &    47.810    &    32.842 &   2.227    &    2.054 &   0.537   &    0.007 &   0.192   &    0.009 &   0.244   &    0.012      \\ 
  
\hline           
\end{tabular}

\end{minipage}}

\end{sidewaystable}


%
%
%
%
%
%
%
%
%
%
%
%

\begin{table}
\caption{PFS Doppler measurements of TOI-2202.} 
\label{table:TIC358107516_PFS} 

\centering  

\begin{tabular}{c c c } 

\hline\hline    
\noalign{\vskip 0.5mm}

Epoch [JD] & RV [m\,s$^{-1}$] & $\sigma_{RV}$ [m\,s$^{-1}$]   \\  

\hline     
\noalign{\vskip 0.5mm}

2458682.9274  & 	   3.29 &   2.40	 \\
2458741.8941  & 	   8.39	 &  2.44   \\
2458761.8289  & 	 $-$84.36 &  2.43	 \\
2458828.7015  & 	$-$122.43 & 	  2.11 \\
\hline           
\end{tabular}
\end{table}

\begin{table*}[ht]

\centering   
\caption{{Continued from \autoref{NS_params}. Nested sampling posteriors and maximum $-\ln\mathcal{L}$ nuisance parameters estimates of the two-planet system TOI-2202 derived by 
joint dynamical modeling of photometry ($\tess$, CHAT) and radial velocities (FEROS, PFS, HARPS). 
}}
\label{NS_params2}

 \begin{adjustwidth}{-1.8cm}{}
 \resizebox{0.85\textheight}{!}
 {\begin{minipage}{1.1\textwidth}

\begin{tabular}{lrrrrrrrrrrrr}     

\hline\hline  \noalign{\vskip 0.7mm}

\makebox[0.1\textwidth][l]{\hspace{0 mm} Parameter  \hspace{35 mm}  Median and $1\sigma$  \hspace{15 mm} Max. $-\ln\mathcal{L}$     \hspace{25 mm} Adopted priors  \hspace{0 mm} \hspace{1.5 mm} } \\
\hline

 RV GP$_{\rm Rot.}$ Amp. [m$^{2}$\,s$^{-2}$]   &  &  585.3$_{-244.3}^{+229.1}$  &  &  615.1   &  &  $\mathcal{U}$(5.0,1000.0)   \\ \noalign{\vskip 0.9mm} 
 
 Transit GP$_{\rm Rot.}$ Amp. [ppm$^{2}$]   &  &  576.0$_{-439.0}^{+1135.0}$  &  &  182.1   &  &  $\mathcal{J}$(100.0,2000.0)   \\ \noalign{\vskip 0.9mm} 
 
 Transit GP$_{\rm Rot.}$ timescale [day]    &  &  150.8$_{-31.2}^{+23.4}$  &  &  201.0   &  &  $\mathcal{J}$(100.0,500.0)   \\ \noalign{\vskip 0.9mm} 
 
 Transit GP$_{\rm Rot.}$ Period [day]     &  &  24.1$_{-1.8}^{+2.3}$  &  &  24.3  &  &  $\mathcal{U}$(20.0,30.0)   \\ \noalign{\vskip 0.9mm} 
  
 Transit GP$_{\rm Rot.}$ fact.         &  &  0.0059$_{-0.0053}^{+0.0353}$  &  &  0.0044  &  &  $\mathcal{J}$(0.0001,0.2)   \\ \noalign{\vskip 1.9mm}

RV offset$_{\rm FEROS}$ [m\,s$^{-1}$] & \hspace{10 mm}  &  --119.5$_{-11.3}^{+10.9}$  & \hspace{20 mm} 
                                      &  --120.6   & \hspace{20 mm} 
                                      &  $\mathcal{U}$(--140.00,--100.00)    \\ \noalign{\vskip 0.9mm} 
RV jitter$_{\rm FEROS}$ [m\,s$^{-1}$] & &  19.1$_{-15.2}^{+10.7}$  &  
                                      &  26.8   &  
                                      &  $\mathcal{J}$(0.00,50.00)    \\ \noalign{\vskip 0.9mm} 
 
RV offset$_{\rm PFS}$ [m\,s$^{-1}$]   & &  --90.1$_{-9.8}^{+10.7}$  &  
                                      &  --98.3  &  
                                      &  $\mathcal{U}$(--110.00,--70.00)    \\ \noalign{\vskip 0.9mm} 
RV jitter$_{\rm PFS}$ [m\,s$^{-1}$]   & &  5.0$_{-3.2}^{+8.7}$  &  
                                      &  3.4   &  
                                      &  $\mathcal{J}$(0.00,50.00)    \\ \noalign{\vskip 0.9mm} 

RV offset$_{\rm HARPS}$ [m\,s$^{-1}$] & &  --122.1$_{-9.4}^{+11.4}$  &  
                                      &  --123.3   &  
                                      &  $\mathcal{U}$(--140.00,--100.00)    \\ \noalign{\vskip 0.9mm} 
RV jitter$_{\rm HARPS}$ [m\,s$^{-1}$] & &  9.8$_{-7.4}^{+8.2}$  &  
                                      &  15.6   &  
                                      &  $\mathcal{J}$(0.00,50.00)     \\ \noalign{\vskip 0.9mm} 
                                      
Transit offset$_{\rm TESS-S1}$ [ppm]  &  &  489$_{-1336}^{+1088}$  &  &  1695  &  &  $\mathcal{U}$(--2000.0,2000.0)     \\ \noalign{\vskip 0.9mm} 
Transit jitter$_{\rm TESS-S1}$  [ppm] &  &  284$_{-233}^{+220}$  &  &  431 &  &  $\mathcal{J}$(0.0,1000.0) \\ \noalign{\vskip 0.9mm} 

Transit offset$_{\rm TESS-S2}$ [ppm]   & &  -348$_{-1229}^{+1372}$  &  &  --1755  &  &  $\mathcal{U}$(--2000.0,2000.0)   \\ \noalign{\vskip 0.9mm} 
Transit jitter$_{\rm TESS-S2}$  [ppm]  &   &  99$_{-72}^{+187}$ &   &  226 &  &  $\mathcal{J}$(0.0,1000.0) \\ \noalign{\vskip 0.9mm} 

Transit offset$_{\rm TESS-S6}$ [ppm]   & &  -261$_{-972}^{+1230}$  &  &  193  &  &  $\mathcal{U}$(--2000.0,2000.0)    \\ \noalign{\vskip 0.9mm} 
Transit jitter$_{\rm TESS-S6}$  [ppm]  & &  85$_{-56}^{+226}$  &  &   48 &  &  $\mathcal{J}$(0.0,1000.0) \\ \noalign{\vskip 0.9mm} 
 
Transit offset$_{\rm TESS-S9}$ [ppm]   & &  -61$_{-1144}^{+1185}$  &  &  --1256  &  &  $\mathcal{U}$(--2000.0,2000.0)   \\ \noalign{\vskip 0.9mm} 
Transit jitter$_{\rm TESS-S9}$  [ppm]   &  &  100$_{-70}^{+239}$  &  &  161 &  &  $\mathcal{J}$(0.0,1000.0) \\ \noalign{\vskip 0.9mm}  

Transit offset$_{\rm TESS-S13}$ [ppm]   & &  4619$_{-3510}^{+2852}$  &  &  3039  &  &  $\mathcal{U}$(--2000.0,2000.0)   \\ \noalign{\vskip 0.9mm} 
Transit jitter$_{\rm TESS-S13}$  [ppm]  &   &  56$_{-34}^{+140}$ &   &  50 &  &  $\mathcal{J}$(0.0,1000.0) \\ \noalign{\vskip 0.9mm}  
 
Transit offset$_{\rm CHAT-1}$ [ppm]   &  &  690$_{-1714}^{+4946}$  &  &  260    &  &  $\mathcal{U}$(--10000.0,10000.0)    \\ \noalign{\vskip 0.9mm} 
Transit jitter$_{\rm CHAT-1}$  [ppm]  &  &  1688$_{-740}^{+1236}$ &   &  878 &  &  $\mathcal{J}$(0.0,5000.0) \\ \noalign{\vskip 0.9mm} 

Transit offset$_{\rm CHAT-2}$ [ppm]  &   &  -1419$_{-1864}^{+3316}$  &  &  --1436    &  &  $\mathcal{U}$(--10000.0,10000.0)    \\ \noalign{\vskip 0.9mm} 
Transit jitter$_{\rm CHAT-2}$  [ppm] &   &  2289$_{-872}^{+1968}$  &  &  1333 &  &  $\mathcal{J}$(0.0,5000.0) \\ \noalign{\vskip 0.9mm} 

Transit offset$_{\rm CHAT-3}$ [ppm]   &  &  4805$_{-2851}^{+2045}$  &  &  4712    &  & $\mathcal{U}$(--10000.0,10000.0)  \\ \noalign{\vskip 0.9mm} 
Transit jitter$_{\rm CHAT-3}$  [ppm] &   &  1674$_{-1048}^{+1841}$  &  &  722 &  &  $\mathcal{J}$(0.0,5000.0) \\ \noalign{\vskip 0.9mm}

Quad. limb-dark.$_{TESS}$ $u_1$  &  &  0.49$_{-0.23}^{+0.19}$  &  &  0.63   &  &  $\mathcal{U}$(0.00,1.00)   \\ \noalign{\vskip 0.9mm}
Quad. limb-dark.$_{TESS}$ $u_2$ &  &  0.44$_{-0.25}^{+0.26}$  &  &  0.05   &  &  $\mathcal{U}$(0.00,1.00)   \\ \noalign{\vskip 0.9mm} 

Quad. limb-dark.$_{\rm CHAT}$ $u_1$  &  &  0.47$_{-0.23}^{+0.25}$  &  &  0.22   &  &  $\mathcal{U}$(0.00,1.00)    \\ \noalign{\vskip 0.9mm} 
Quad. limb-dark.$_{\rm CHAT}$ $u_2$  &  &  0.54$_{-0.28}^{+0.26}$  &  &  0.89   &  &  $\mathcal{U}$(0.00,1.00)   \\ \noalign{\vskip 0.9mm}

\\
\hline \noalign{\vskip 0.7mm}

\end{tabular}

\end{minipage}}
\end{adjustwidth}

\end{table*}

\end{document}